\documentclass[fleqn,usenatbib]{mnras}

\usepackage{newtxtext,newtxmath}

\usepackage[T1]{fontenc}

\usepackage{color}

\DeclareRobustCommand{\VAN}[3]{#2}
\let\VANthebibliography\thebibliography
\def\thebibliography{\DeclareRobustCommand{\VAN}[3]{##3}\VANthebibliography}

\usepackage{graphicx}	
\usepackage{amsmath}	

\usepackage{bm}
\usepackage{booktabs}
\usepackage{multirow}

\usepackage{subfig}

\title[Biases from differential magnification]{Biases in galaxy spectral analysis from strong lensing differential magnification effect and correction methods}

\author[Xu et al.]{
Wenshuo Xu$^{1}$\thanks{E-mail: \url{xws21@mails.tsinghua.edu.cn}},
Dandan Xu$^{1}$\thanks{E-mail: \url{dandanxu@tsinghua.edu.cn}},
Xinzhong Er$^{2}$
and Junqiang Ge$^{3}$
\\
$^{1}$Department of Astronomy, Tsinghua University, Beijing 100084, China\\
$^{2}$South-Western Institute for Astronomy Research, Yunnan University, Kunming, 650500, China\\
$^{3}$National Astronomical Observatories, Chinese Academy of Sciences, 20A Datun Road, Chaoyang District, Beijing 100101, China
}

\date{Accepted XXX. Received YYY; in original form ZZZ}

\pubyear{2024}

\begin{document}
\raggedbottom
\label{firstpage}
\pagerange{\pageref{firstpage}--\pageref{lastpage}}
\maketitle

\begin{abstract}
Strong gravitational lensing has significantly advanced the study of high-redshift galaxies, but the differential magnification effect inevitably introduces biases in the spectral analysis of source galaxies.
This work investigates these biases using mock lensing systems from MaNGA survey data and IllustrisTNG simulations. We analyze the impact of lensing effect on several spectral properties, including stellar age, metallicity, H$\alpha$ flux, and optical emission line ratios. Our results show significant biases in all properties after lensing. The values of quantities can be either over- or under-estimated, except for the consistently enhanced H$\alpha$ flux. The bias varies with lensing configurations and always arises when part of the source galaxy falls into the strong lensing regime. We evaluate two correction methods to recover the intrinsic source properties: the average magnification factor ($\bar{\mu}$) and full ray-tracing. While both methods reduce the overestimated H$\alpha$ flux, the $\bar{\mu}$ method shows a much larger discrepancy. For stellar population properties and emission line ratios, the $\bar{\mu}$ method fails whereas the ray-tracing method proves effective. Applying these two methods to a statistical sample of mock systems further shows their strong dependence on lens modeling accuracy. As a demonstrative study, our results highlight the importance of spatially resolved spectroscopic observations and precise lens modeling for reconstructing spectra of strongly lensed galaxies. While our conclusions are based on a specific source and lens galaxy, further studies with a statistical sample of realistic mock lensing systems are needed for understanding any systematic differences between the two correction methods.

\end{abstract}

\begin{keywords}
gravitational lensing: strong -- galaxies: high-redshift
\end{keywords}


\section{Introduction}
\label{sec1:introduction}
The study of high-redshift galaxies has imposed crucial constraints on the formation and evolution of galaxy and cosmology \citep{Santini2009, Stark2016, Girelli2019, Carnall2023, Weibel2024, Haslbauer2022, Boylan2023}. Over the past few decades, galaxy surveys have revealed a clear history of the cosmic star-formation, which peaked at $z\sim2.0$ \citep{CosmicNoon}. Further powerful ground-based facilities (e.g. ALMA, VLT) and space-borne telescopes (e.g. HST, JWST, \textit{Euclid}) have made a leap forward in the study of early galaxies and push the detection limit to $z>7$, offering glimpses into universe only a few hundred million years after the Big Bang \citep[e.g.][]{Laporte2021, Robertson2022, Harikane2023, CEERS_paper}.

To characterize high-redshift galaxies, spectral analysis provides a wealth of information. The properties of stellar population, such as stellar metallicity, age and star formation history, are encoded in the continuum and absorption lines, while the emission-line diagnostics reveals the metallicity and ionization state of the gas component \citep{EMLineDiag}. The development in spectroscopic techniques, such as slitless spectroscopy and integrated field spectroscopy, have significantly improved the spatially resolved spectra of individual galaxies, enabling detailed analysis of their interior kinematics and spatial variation of properties \citep{Wylezalek2018, Zhu2023, Lu2023}.

Strong gravitational lensing enhances the discovery and study of high-redshift objects by magnifying their overall flux, enabling high signal-to-noise ratio photometric and spectroscopic measurements \citep{Stark2013, Hashimoto2018, Atek2023}. Additionally, the lensing effect improves the angular resolution of the source plane, which facilitates spatially resolved studies of distant galaxies \citep{Stark2008, Birkin2023, Gim2024}. In current observations, strong lensing systems are rare, with only hundreds of galaxy-scale strong lensing systems being discovered, in which galaxy-sized halos play the role of the lens. However, the predicted number for the ongoing and forthcoming surveys (e.g. LSST, \textit{Euclid}, CSST, \textit{Roman}) can reach a level several orders of magnitude higher than the current one \citep{Collett2015, Weiner20, Cao2024}. In order to study the properties of background source objects, it is necessary to conduct precise lens modeling, which is generally complicated and time-consuming.

Although spatially resolved spectroscopic observations help to study detailed properties of galaxies, it is demanding to conduct such observations for all strong lensing systems and only composite spectra of most lensed galaxies have been obtained at present. The lensing magnification is wavelength independent but can vary rapidly with different source positions. As a result, for strong lensing systems, different parts of the source galaxies are magnified with varying magnification factors \citep{Hezaveh2012, Er2013}. Such differential magnification can introduce biases into the lensed composite spectrum in case the spectral properties of the source galaxy exhibit significant spatial variations, which is well-known for high-redshift galaxies \citep[e.g.][]{Shibuya2016, Wang2020}.

Differential magnification effect has been used to interpret the abnormal spectral properties of lensed high-redshift galaxies, such as the observed SED \citep{Blain1999, Zotti2024} and flux ratios of spectral lines \citep{Downes1995}. Millimeter-wave observation of several strong lensing systems also uncovered the asymmetric double-horned emission line profile due to the differentially magnified rotating source galaxies \citep{Rybak2015, Leung2017, Paraficz2018}. \citet{Serjeant2012} investigated the biases from differential magnification effect with simulated multi-component source and parametric lens and also showed the variations of such biases with different lensing configurations (see also \citealt{Hezaveh2012} and \citealt{Er2013}). \citet{Motta2018} re-derived the property of a high-redshift galaxy lensed by the {\it Bullet Cluster} with a newly discovered image far from the critical line, clarifying the uncertainties in previous results introduced by the differential effect. Furthermore, \citet{Er2013} studied the calibration with average magnification factor and found that the bias still remains after correction.

In this work, we follow the methodology of \citet{Er2013} to study the biases of the differential magnification effect with more realistic mock lensing systems. We generate mock systems with IFU datacubes from the MaNGA survey \citep{Bundy2015} and simulated galaxies in the IllustrisTNG simulations. Then we compare the composite spectral properties of source galaxy before and after being lensed to analyze the biases in stellar population properties, such as stellar metallicity and age, and the emission-line diagnostics. We also generate a suite of mock systems with varying source positions to study the bias in average. Furthermore, we correct the biased composite spectra with average magnification factor and ray-tracing, respectively, to assess the performance of these two correction methods.

This paper is organized as follows. We describe the data used for generating mock lensing systems in Section \ref{sec2:data preparation} and present the methods of ray-tracing simulation, lens modeling and source reconstruction in Section \ref{sec3:Methods}. Section \ref{sec4:composite spectrum after lensed} discusses the bias in source spectral analysis from differential magnification effect and its variation with different lensing configurations. We apply the two correction methods and compare their performance in Section \ref{sec5:correction for biases}. Finally, a brief discussion and summary of our results is given in Section \ref{sec6:summary}. In this work, we adopt the cosmology used in IllustrisTNG simulation from \citet{Planck2016} with cosmological parameters: $\Omega_m=0.3089$, $\Omega_{\Lambda}=0.6911$ and $h=0.6774$.



\begin{figure}
\centering
\includegraphics[width=0.33\textwidth]{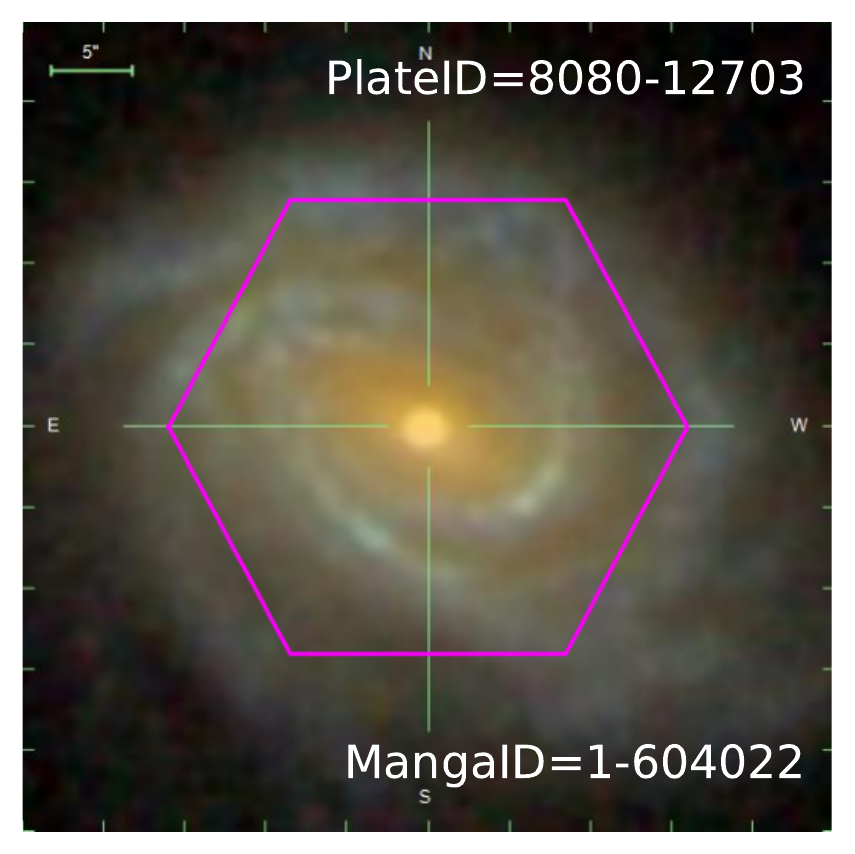}
\caption{The SDSS three-color RGB image of the selected source galaxy in this work. The overlaid magenta hexagonal box indicates the footprint of the MaNGA IFU. The corresponding plateID and MangaID are annotated in the figure.}
\label{Fig1_selected_samples}
\end{figure}

\section{Data Preparation}
\label{sec2:data preparation}

In this work, we aim to investigate biases in source spectral analysis induced by strong lensing effect with mock lensing systems. In this section, we present the galaxies that we select and use as the background source and foreground lens in the mock lensing systems. We select a galaxy from the MaNGA survey and place it at $z=2.0$ as the source galaxy (Section \ref{sec2.1:MaNGA Galaxies}), and a simulated galaxy from snapshot $z=0.5$ in the TNG-100 simulation as the foreground lens galaxy (Section \ref{sec2.2:TNG data}).

\subsection{Source: a MaNGA galaxy}
\label{sec2.1:MaNGA Galaxies}

\textit{Mapping Nearby Galaxies at Apache Point Observatory} (MaNGA) is one of the core programs in SDSS-IV, aiming at obtaining spatially resolved spectroscopic data for about 10,000 nearby galaxies, in order to investigate detailed galaxy properties such as stellar and gas kinematics, stellar populations and so on \citep[see][]{Bundy2015}.  MaNGA employs 17 fiber-bundle Integral Field Units (IFUs) with a fiber number varying from 19 to 127, providing two-dimensional spatially resolved spectroscopic data between 3600 \AA\  and 10300 \AA\ with a spectral resolution of $R\sim2000$.

MaNGA Data Reduction Pipeline \citep[DRP,][]{Law2016, Law2021} produces datacubes at a spatial resolution of 0.5 arcsec/pixel. In this study, we take a MaNGA galaxy in SDSS-IV DR17 \citep{mangadata} as the source galaxy to generate the mock spectrum distribution on the image plane and study the effect of differential magnification, which demands its IFU observation to be sufficiently resolved. To this end, we first select galaxy candidates observed with 127 fibers in the IFU plate, which shall largely cover the luminous radii in their SDSS RGB images, as shown for the four example galaxies in Fig.\,\ref{Fig1_selected_samples}. Among these galaxies, we further check their spatially-resolved properties based on the results from MaNGA DynPop \citep{Lu2023}. Ultimately, we select one galaxy (MaNGA plateID: 8080-12703, also named as NGC 1280, the first galaxy shown in Fig.\,\ref{Fig1_selected_samples}), which exhibits large radial gradients in both stellar age and metallicity distributions \citep[$\gamma_{\rm{age}}=-1.23$, $\gamma_{Z_*}=-0.47$, see definitions in][]{Lu2023}{}{}, as the source to generate our mock lensing systems. In addition, this galaxy has also been confirmed to host an AGN at its centre in previous studies \citep[e.g.][]{Comerford2020, Chen2022}. As presented in the following sections, the composite spectra of such galaxies with significant intrinsic variations of spectral properties can be markedly biased after being lensed.

Here we utilize two types of observables of this galaxy. One is the V-band surface brightness distribution, which is used to generate mock images after being gravitationally lensed by a foreground galaxy. Here the intrinsic V-band image is obtained from the IFU datacube by combining luminosities of voxels within this waveband. The other is the spatially resolved spectrum distribution, which is used to generate mock IFU datacubes on the image plane. The MaNGA galaxy is located at $z=2.0$ as our mock source, and its spectrum is redshifted, the brightness is decreased, and the angular size is re-scaled accordingly.

\begin{figure*}\centering
	\includegraphics[width=\textwidth]{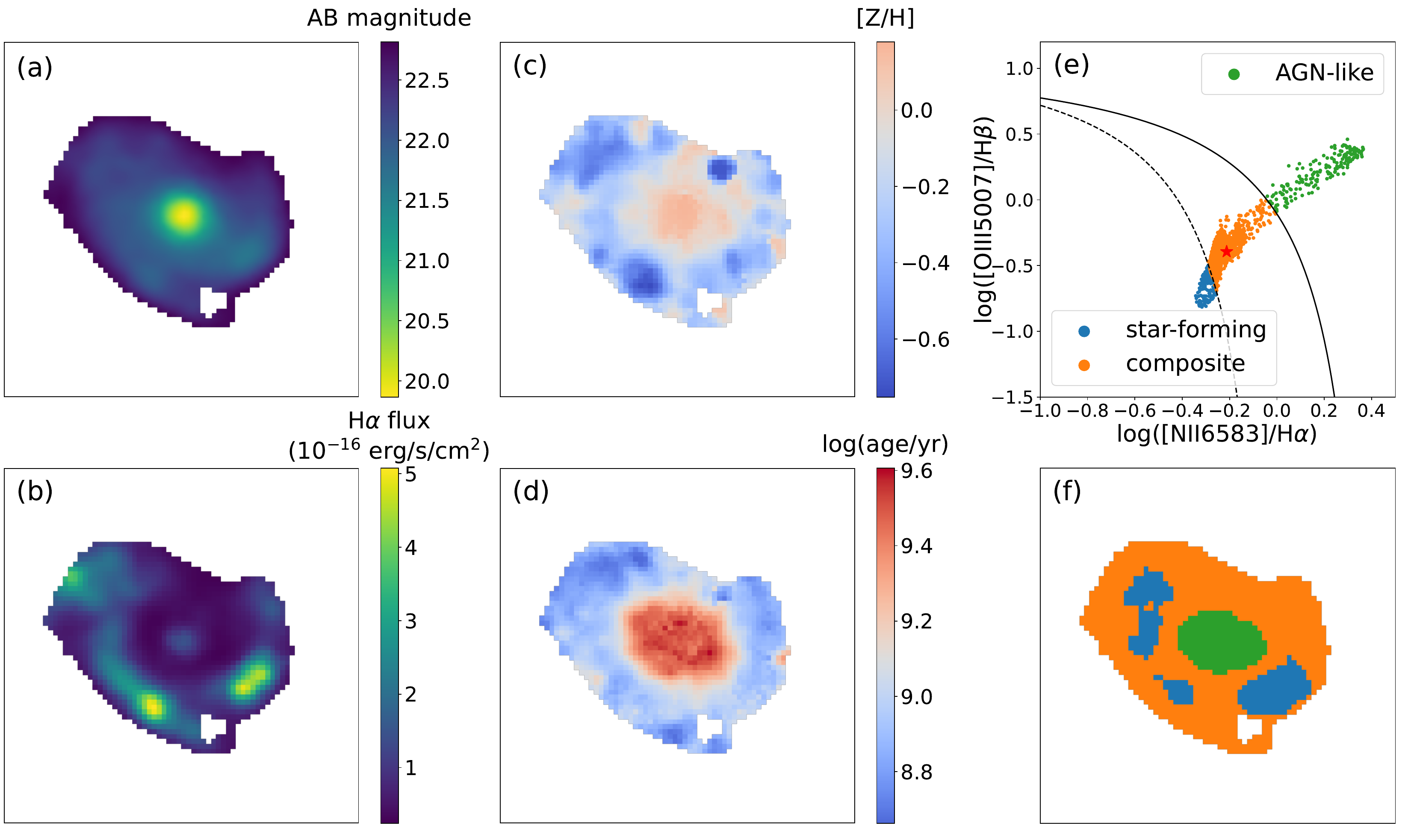}
    \caption{V-band image of source galaxy (MaNGA plateID: 8080-12703) and internal spatial distributions of spectral properties. Panel (a) shows the V-band image generated from MaNGA IFU datacube as discussed in Section \ref{sec2.1:MaNGA Galaxies}. Panel (b) shows the H$\alpha$ flux map. Panel (c) and (d) show the spatial variations of stellar metallicity and age within this galaxy, respectively. Here we adjust the colorbar to plot regions with stellar metallicity/age higher than the composite values (shown in Tab.\,\ref{tab:fitted_param}) in red and regions with lower values in blue. Panel (e) is the spatially resolved N2-BPT diagram. The dashed \citep{Kauffmann2003} and solid lines \citep{Kewlwy2001} show the empirical and theoretical separations of star-forming galaxies, composites and AGNs. Corresponding regions are marked with the same color in Panel (f). The red star in Panel (e) indicates the location of galaxy's composite spectrum.}
    \label{Fig2_src_image}
\end{figure*}

With such source information, we use the ray-tracing method presented in the Section \ref{sec3.1: mock lensing systems} to generate the mock lensed image and spectrum distribution on image plane. However, directly using observational data to generate mock lensing systems can be problematic due to the observational effects and pre-processing of the data is necessary. Firstly, voxels with low S/N ratio can lead to biases in the following spectral analysis and need to be masked. As the S/N ratio decreases with lower surface brightness under the same exposure time, here we fit the galaxy V-band image with an elliptical S{\'e}rsic model \citep{Sersic_profile} and artificially mask pixels with surface brightness lower than $0.5\,I_0$ and the related voxels, where $I_0$ is the best-fit S{\'e}rsic surface brightness at the effective radius. Secondly, we linearly interpolate the original datacube onto a mesh of higher spatial resolution to relieve the limit of MaNGA resolution. The V-band image after mask and interpolation is shown in Panel (a) of Fig.\,\ref{Fig2_src_image}. We have checked that such operations do not deviate our final conclusions. The FWHM of instrinsic MaNGA PSF varies between 1.0 to 1.5 arcsec in multiple observations and becomes 0.05 to 0.08 arcsec after rescaled to $z=2.0$, which is comparable to HST resolution and much smaller than the effective radius of source galaxy ($\sim0.5$ arcsec at $z=2.0$). Since both the biases and subsequent lens modeling process in this work mostly rely on the overall profile of source galaxy, here we ignore the intrinsic PSF effect in MaNGA data.

We study the spatial distribution of different spectral properties within this galaxy using the \textsc{ppxf} software \citep{Cappellari2004,Cappellari2017} and the Vazdekis/MILES stellar population library \citep{Vazdekis2010} for spectral fittings \citep{Ge2018,Ge2019}.
We truncate the spectrum of the galaxy at 8622\AA\ due to the significant noise at longer wavelength. The spatial distribution of H$\alpha$ flux, luminosity-weighted stellar metallicity, stellar age and spatially resolved N2-BPT diagram of the source galaxy are shown in Fig.\,\ref{Fig2_src_image}. The fitting result shows that the central region of this galaxy is of higher stellar metallicity and age, which is consistent with \citet{Lu2023}. The spatially resolved N2-BPT diagram indicates a central AGN and a star-forming disc with clumpy star-formation regions showing strong H$\alpha$ emission. We have checked that our emission line maps consist with the results of MaNGA Data Analysis Pipeline \citep[DAP,][]{mangaDAP}.

\subsection{Lens: a Simulated Galaxy in TNG-100}
\label{sec2.2:TNG data}

\textit{The Next Generation Illustris Simulations} \citep[TNG hereafter,][]{TNG_1, TNG_2, TNG_3, TNG_4, TNG_5, TNG_6, TNG_7} are a series of magnetohydrodynamic cosmological galaxy formation simulations carried out with the moving mesh code \textsc{arepo}. As a successor to the hydrodynamic simulation \textit{Illustris}, TNG includes the effect of magnetic field and new AGN feedback models and optimizes the original galaxy formation and evolution models. TNG project consists of three primary runs with different box sizes and resolutions. In this work, we use the full-physics version TNG-100 simulation with box size 75 Mpc/$h$ and mass resolution $\sim 10^6 M_{\odot}$. Following the fact that elliptical galaxies always have larger lensing cross sections than discs, we select a massive elliptical in snapshot 067 ($z=0.5$) as the lens in the mock system.

The selected galaxy has a {\sc subfind} ID of 378918 at snapshot 067.
It has a total mass of $1.3\times 10^{12} M_\odot $, a half total mass radius of 67.2 kpc and contains 153,292 dark matter particles, 47,446 stellar particles and 80,572 gas elements. We note that the selected galaxy is the central galaxy of a galaxy-scale dark matter halo and is not part of a group- or cluster-environment. We take the $z$-projection of the galaxy to generate the mass distribution in the lens plane. Here we only use particles belonging to the main halo and exclude all other subhalos and substructures within it. The Einstein radius is estimated as the radius within which the enclosed mean surface density is equal to the critical surface mass density $\Sigma_{\rm cr}$ (Eq.\ref{eq:convergence}) given the adopted lens and source redshifts for the mock system \citep{Narayan1996}.  
Along this projection, the estimated Einstein radius $\theta_{\rm E}$ is $0.55^{\prime\prime}$, corresponding to 3.46 kpc at $z=0.5$, which is well consistent with the derived Einstein radius from fitting a singular isothermal ellipsoid (SIE) as the lens to the mock image observation (see Section \ref{sec3.1: mock lensing systems}). The axis ratios of the luminous (stellar) and the total matter components are calculated iteratively within an elliptical aperture using the second-moment method (see \citealt{Xu2017} for details). The luminous axis radio $q_*$ and the matter axis ratio $q$ measured within a radius of $3\theta_{\rm E}$, are 0.63 and 0.87, respectively. In Section \ref{sec3.1: mock lensing systems}, we introduce the numerical method that we calculate lensing properties of the simulated galaxy.


\section{Methodology}
\label{sec3:Methods}

In this section, we present our methods to generate mock lensing systems in Section \ref{sec3.1: mock lensing systems}, including a brief review of strong lensing formalism (Section \ref{sec3.1.1: strong lensing formalism}), a summary to the methods that we use to calculate lensing potential and its first and second order derivatives of the lens galaxy (Section \ref{sec3.1.2: AlphaMuMap}), to generate mock images and spectrum datacube in the image plane (Section \ref{sec3.1.3:raytracing method}), to add observational effect (Section \ref{sec3.1.4:obs_effect}) and finally to generate a statistical sample of mock systems (Section \ref{sec3.1.5:systems_in_this_work}). We then present the method that we use to fit lensing models to the mock systems in Section \ref{sec3.2:lens_modeling}. Finally two different approaches to reconstruct the source spectrum are introduced in Section \ref{sec3.3:src_rec}.

\subsection{Mock Lensing Systems}
\label{sec3.1: mock lensing systems}

\subsubsection{Basic strong lensing formalism}
\label{sec3.1.1: strong lensing formalism}

Throughout this work, we adopt thin-lens approximation that the deflection effect of the foreground lens only depends on the redshifts of the lens and the source, and the projected mass distribution $\Sigma(\vec{\theta})$ in the lens plane perpendicular to the line of sight. Consistent with the notations in \citet{Narayan1996}, here we denote angular coordinates in the source plane as $\vec{\beta}$ and in the image plane as $\vec{\theta}$. Then the lens equation describing light deflection is written as:
\begin{equation}
\vec{\beta}=\vec{\theta}-\vec{\alpha}(\vec{\theta}),
\label{eq:lenseq}
\end{equation}
where $\vec{\alpha}(\vec{\theta})$ is the reduced deflection angle as a function of $\vec{\theta}$. With thin-lens approximation, the deflection angle $\vec{\alpha}(\vec{\theta})$ is the first derivative of the effective lensing potential $\psi(\vec{\theta})$, which is a normalized and integrated 3D gravitational potential along the sight line to the lens, defined as: 
\begin{equation}
\psi(\vec{\theta})=\frac{D_{\rm{ds}}}{D_{\rm{d}}D_{\rm{s}}}\frac{2}{c^2}\int \Phi(D_{\rm{d}}\vec{\theta}, z)\mathrm{d}z=\frac{1}{\pi} \int \kappa(\vec{\theta^{\prime}}) \ln{\lvert\vec{\theta}-\vec{\theta^{\prime}} \rvert}\ \mathrm{d}^2 \vec{\theta^{\prime}}
\end{equation}
where $\kappa(\vec{\theta})$ is the normalized surface density $\Sigma(\vec{\theta})$ of the lens matter distribution, called the \textit{convergence}
\begin{equation}
\label{eq:convergence}
\kappa(\vec{\theta}) = \frac{\Sigma(\vec{\theta})}{\Sigma_{\rm{cr}}},\quad \Sigma_{\rm{cr}}=\frac{c^2}{4\pi G} \frac{D_{\rm{s}}}{D_{\rm{d}} D_{\rm{ds}}},
\end{equation}
where $\Sigma_{\rm{cr}}$ is the critical surface density and $D_{\rm{s}}$, $D_{\rm{d}}$ and $D_{\rm{ds}}$ are the angular diameter distances between observer and source, observer and lens and lens and source, respectively. 
 
The deflection angle $\vec{\alpha}(\vec{\theta})$ then becomes the convolution of $\kappa(\vec{\theta})$ with the 2D kernel $\vec{\theta}/\lvert \vec{\theta} \rvert^2$,
\begin{equation}
\label{eq:deflection angle}
    \vec{\alpha}(\vec{\theta})=\vec{\nabla} \psi(\vec{\theta})=\frac{1}{\pi} \int \kappa(\vec{\theta^{\prime}}) \frac{\vec{\theta} - \vec{\theta^{\prime}}}{\lvert \vec{\theta} - \vec{\theta^{\prime}} \rvert^2}\ \mathrm{d}^2 \vec{\theta^{\prime}}. 
\end{equation}
The deflection can cause convergence to the light, i.e. the magnification effect. For a point source, the magnification factor $\mu(\vec{\theta})$ can be calculated from the derivative of Jacobian matrix $\mathcal{A}=\partial\vec{\beta}/\partial\vec{\theta}$ \citep[see Eq.53-60 in][]{Narayan1996}.

Positions $\{\vec{\theta}_{\rm{cri}}\}$ in the image plane with $\lvert \mu(\vec{\theta}_{\rm{cri}}) \rvert=\infty$ are named as the \textit{critical curves}.
The corresponding positions in the source plane via Eq.\,\ref{eq:lenseq} are the \textit{caustics}. The magnification magnitude increases when a source approaches to the caustics. In the case of an extended source, different parts can be magnified with different magnification factor, which is called \textit{differential magnification effect}. In this case, the magnification will not be infinite and an average magnification $\bar{\mu}$ is defined as the ratio of the total flux after and before it being lensed: 
\begin{equation}
\label{eq:ave_mag}
    \bar{\mu}=\frac{\int I(\vec{\theta}) \mathrm{d}^2 \vec{\theta}}{\int I^{\rm s} (\vec{\beta}) \mathrm{d}^2 \vec{\beta}},
\end{equation}
where $I(\vec{\theta})$ and $I^{\rm s}(\vec{\beta})$ are the surface brightness of the image and the source on the image and the source plane, respectively. Since gravitational lensing preserves the surface brightness, given by the Liouville's theorem \citep{Book_Schneider}, the two surface brightness distributions are linked by $I(\vec{\theta})=I^{\rm s}[\vec{\beta}(\vec{\theta})]=I^{\rm s}[\vec{\theta}-\vec{\alpha}(\vec{\theta})]$.

\subsubsection{Deflection angle and magnification maps}
\label{sec3.1.2: AlphaMuMap}

In this work, we employ meshes to calculate the lensing potential $\psi(\vec{\theta})$ using the Particle-Mesh method with an implementation of FFT to speed up computation. Once the potential field is obtained, the first- and second-order derivatives of the lensing potential are then derived using a five-point stencil finite difference method. We mainly follow the procedure presented in \citet{Xu2009}. Here we recap the key ingredients in this calculation.

We shall note that in addition to the local surface density, matter distributing further out also influences the deflection angle distribution in the central strong lensing domain of few kiloparsecs via providing shear, as can be seen from the infinite convolution kernel in Eq.\ref{eq:deflection angle}. For this reason, we need a mesh to cover a fairly large region of the simulated galaxy halo, such that exclusion of any matter distribution beyond would not significantly affect the result of the investigation here and thus can be safely ignored. This naturally requests a large grid dimension to reach sufficient spatial resolution in the central few kpc region for strong lensing, which will however significantly increase the computational expense.

To guarantee sufficient spatial resolution while saving computational power, we adopt a set of two-level nested meshes in the lens plane. One covers the central region to capture the local convergence in the strong lensing domain, while the other one covers a much larger region to correctly pick up the effect of shear from matter further out. For the lens galaxy used in this work, both meshes have a dimension of $2048\times 2048$. Particles of the simulated galaxy are assigned to the meshes via an SPH scheme \citep{SPHkernel} to obtain the smoothed convergence distributions $\kappa(\vec{\theta})$ in two different domains and with two different spatial resolutions. Since an isolated boundary condition is adopted through a zero-padding technique \citep{Hockney1988}, the mesh sizes need to be two times the size of the matter domain in each case. In this way, the central finer mesh that covers a central squared region of (4.7 arcsec)$^2$ has a side-length of 9.42 arcsec with a spatial resolution of 
$< 0.01\,\theta_{\rm E}$/pixel. Then we truncate the simulated galaxy halo at a radius of twice the half mass radius from the centre and therefore the outer coarse mesh is adopted to cover a spherical matter distribution within a radius of 134.4 kpc and has a size-length of 537.6 kpc with a spatial resolution of 
$< 0.1\,\theta_{\rm E}$/pixel. Once the potential field as well its first- and second-order derivatives are obtained, the results are finally interpolated on to the image plane, which is covered by another mesh with a dimension of $1290 \times 1290$ covering the central strong lensing region of (3.54 arcsec)$^2$ with a finer resolution of $\sim 0.005\,\theta_{\rm E}$/pixel.

\begin{figure*}\centering
	\includegraphics[width=0.75\textwidth]{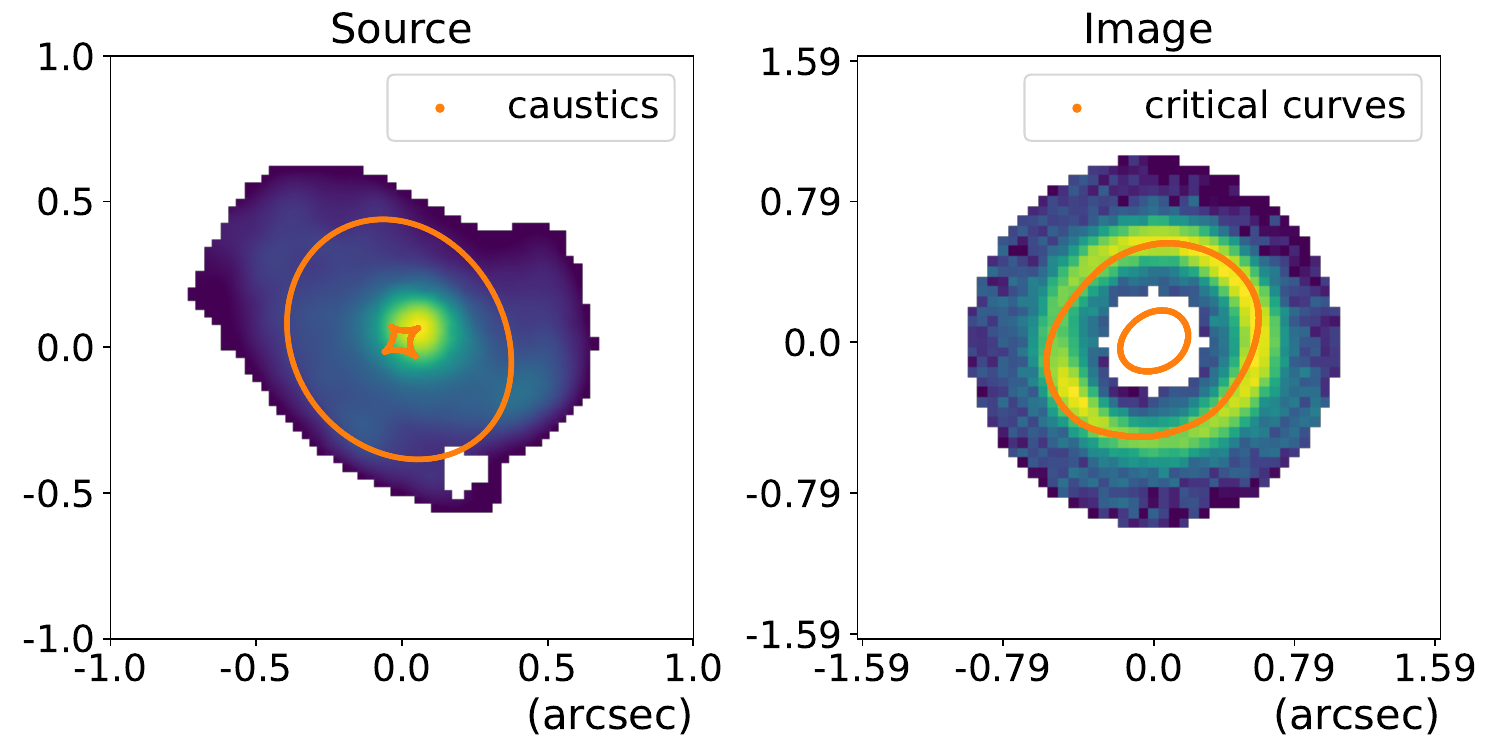}
    \caption{\textit{Left Panel:} Source image adapted to $z=2.0$. \textit{Right Panel:} Image after being lensed. Here we only leave the ring region used for lens modeling. We choose the effective radius of the lens galaxy (0.28 arcsec for the lens galaxy here at $z=0.5$, derived from simulation data) as the inner radius to avoid the possible light contamination of the foreground lens in real observations. The outer radius is artificially set as 1.0 arcsec to simulate a brightness threshold. The orange curves in the two panels show the lens caustics and critical curves, respectively. These two panels share the same colorbar, which is omitted here.}
    \label{Fig3_lensed_image}
\end{figure*}

\begin{figure}
\centering
\includegraphics[width=0.5\textwidth]{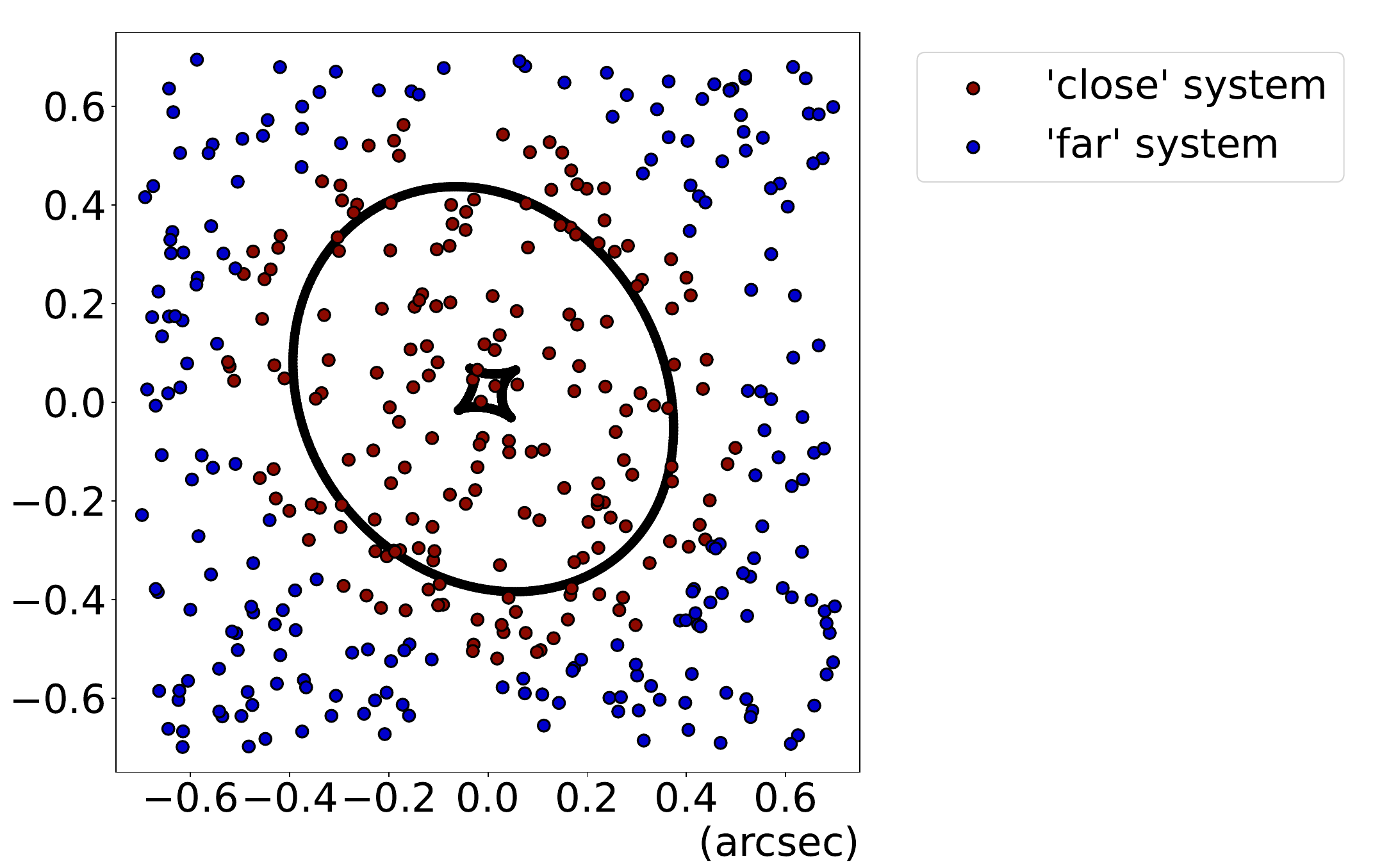}
\caption{The 400 mock lensing systems for the statistical study. Points indicate the relative positions of source light center to the lens caustics in each system. Different colours are used to show the `close' (red) and `far' (blue) subgroups in Section \ref{sec3.1.5:systems_in_this_work}.}
\label{Fig4_400systems}
\end{figure}

\subsubsection{Ray-tracing method}
\label{sec3.1.3:raytracing method}

With deflection angle map $\vec{\alpha}(\vec{\theta})$, we can trace image plane positions $\{\vec{\theta}\}$ back to source plane to get $\{\vec{\beta}(\vec{\theta})\}$ following Eq.\,\ref{eq:lenseq}. Then the surface brightness distribution in the image plane is simply $I(\vec{\theta})=I^{\rm s}[\vec{\beta}(\vec{\theta})]=I^{\rm s}[\vec{\theta}-\vec{\alpha}(\vec{\theta})]$ since gravitational lensing effect preserves surface brightness. This is the \textit{ray-tracing} method and widely used to generate mock strongly lensed images.

To do so, we loop over all mesh points in the image plane, and for any given mesh point $\vec{\theta}_i$, we derive the corresponding source position $\vec{\beta}_i$ via Eq.\ref{eq:lenseq} and assign the surface brightness $I^{\rm s}(\vec{\beta_i})$ to the grid at $\vec{\theta}_i$. According to the achromatic nature of gravitational lensing, the same deflection angle map can be used for ray-tracing at all wavelengths. We note that in order to accurately generate lensed images and spatially resolved spectrum distribution on the $1290 \times 1290$ image mesh, we linearly interpolate the redshift-rescaled MaNGA datacube to a finer grid with 50 times higher spatial resolution. This is necessary to guarantee accurate mock observations in the image plane (see Section \ref{sec2.1:MaNGA Galaxies}).

\subsubsection{Observational effects}
\label{sec3.1.4:obs_effect}

To simulate real observations, we further take into account the observational effects including limited spatial resolution in image plane, PSF effect and noise. For the mock images, here we adopt the instrument parameters of \textit{Hubble Space Telescope} (HST) WFC3, whose PSF FWHM is $\sim$ 0.07 arcsec and spatial resolution is $\sim$ 0.04 arcsec in optical waveband \citep{HST_Handbook}. We convolve the generated images with a Gaussian PSF and then rebin the result down to a resolution of 0.04 arcsec/pixel. Here implicitly ignore the changes of PSF size and spatial resolution with wavebands. Observational noise in data consists of several different parts: the Poisson noise of the target object, the Poisson noise of the sky background, readout noise, and dark current of the instrument. For simplicity, we assume the background-dominated condition and use a standard normal distribution to approximate the Poisson noise of sky background when generating mock lensed images. Specifically, here we adopt an average signal-to-noise ratio S/N of 10. For the mock spectrum datacube, we rebin the mock spectrum generated on the image mesh down to a spatial resolution 0.3 arcsec/pixel to approximate existing MaNGA-like IFU observations but ignore the PSF effect and noise.

\subsubsection{Mock systems in this work}
\label{sec3.1.5:systems_in_this_work}

Following the procedure discussed before, we generate mock lensing systems with a MaNGA galaxy as the source and a TNG galaxy as the lens. Fig.\,\ref{Fig3_lensed_image} shows the source and image plane of one example mock system. In this configuration, the central bulge component of source galaxy is close to the tangential caustic and forms a bright ring image. The mask of lensed image will be discussed more in Section \ref{sec3.2.2:bayesian inference}.

In order to investigate the variation of biases with different lensing configurations, we further generate 400 mock systems by locating the source galaxy at different randomly chosen positions in the source plane to build up a statistical sample. Fig.\,\ref{Fig4_400systems} presents the 400 loactions of source light center in the source plane overlapped with the caustics. The source positions are confined within a certain distance to the caustics so that at least part of the source galaxy is in the strong lensing domain and lensed into multiple images. We further divide these systems into `close' (red) and `far' (blue) subgroups with the criterion whether the central bright bugle is strongly lensed or not. In Section \ref{sec5.2:correction_statistics}, we utilize these two subgroups to study the influence of differential magnification and the performance of source reconstruction in detail.

\subsection{Fitting lens model to mock systems }
\label{sec3.2:lens_modeling}

In this section, we introduce our lens modeling process. We follow the parametric modeling approach commonly used in observational studies, in which the lens and source galaxies are described with parametric models, and the parameters are constrained through Bayesian Inference method based on the lensed image. The best-fit models are adopted in the subsequent corrections. We present the parametric models in Section \ref{sec3.2.1:lens models} and the Bayesian method to fit parameters in Section \ref{sec3.2.2:bayesian inference}.

\subsubsection{Models}
\label{sec3.2.1:lens models}

Modeling of galaxy-galaxy strong lensing systems involves three components: the light profile of the background source, the mass distribution and light profile of the foreground lens. In this work, for simplicity, we ignore the light of the lens galaxy and only model the other two.

Here we fit the mass distribution of lens with the widely used Singular Isothermal Ellipsoidal plus a constant external shear (\textit{SIE + ES}) model. The fact that mass distribution of massive elliptical galaxies can be approximated with a SIE profile \citep{Kormann1994} has been shown in many studies \citep[e.g.]{Treu04, Koopmans06, Auger10, Cappellari13, Oguri14}. The convergence $\kappa(\vec{\theta})$ of the SIE profile is given by:
\begin{equation}
\kappa(x,y)=\frac{\theta_{\rm{E}}}{2\sqrt{q_{\rm l} x^2+y^2/q_{\rm l}}},
\label{eq:SIEmodel}
\end{equation}
where $q_{\rm l}$ is the ratio between the semi-minor and semi-major axes of the elliptical profile. The constant shear component accounts for the gravitational effect of surrounding large-scale structures.

For the source galaxy, we use an elliptical Sérsic profile \citep{Sersic_profile} to model the surface brightness distribution
\begin{equation}
    I(x,y)=I_0 \exp{\left\{-b_n\left[\left(\frac{\sqrt{q_sx^2+y^2/q_s}}{R_{\rm{s}}}\right)^{1/n}-1\right]\right\}},
\label{eq:Sersic}
\end{equation}
where $R_{\rm{s}}$ is the effective radius, $I_0$ is the surface brightness at $R_{\rm{s}}$, $q_s$ is the axis ratio and $n$ is the Sérsic index. The parameter $b_n$ here is approximated as $b_n\approx1.999n-0.327$.

\subsubsection{Bayesian Inference}
\label{sec3.2.2:bayesian inference}

With the parametric models for lens and source, we fit the mock systems in a Bayesian inference way. The posterior probability is written as
\begin{equation}
    P(\bm{\theta}\vert D)=\frac{P(D \vert \bm{\theta})P(\bm{\theta})}{P(D)},
\end{equation}
where $\bm{\theta}$ represents parameters to be fitted, $D$ is the observational data, $P(D\vert\bm{\theta})$ is the likelihood function, $P(\bm{\theta})$ is the prior probability and $P(D)$ is the marginal likelihood. 

In this work, we use a uniform prior and Gaussian likelihood function
\begin{equation}
    \log\mathcal{L}=-\sum_i\left(\frac{f_i-f_{i,\rm{model}}}{\sigma_i}\right)^2,
\label{eq:likelihood}
\end{equation}
where $f_i$ and $f_{i,\rm{model}}$ are the surface brightness of lensed images from mock images and parametric models respectively, and the summation is over all image pixels selected for fitting. The term $\sigma_i$ in the likelihood function is the flux uncertainty. Here we attribute it to be the noise in obervations and determined by the image S/N ratio. As shown in Fig.\,\ref{Fig3_lensed_image}, we further mask the lensed plane into a ring region and only use pixels within it to calculate the likelihood function. We use the effective radius of the foreground lens galaxy as the inner radius of the ring to mask possible light contamination of the lens in real observations. The outer radius is artificially set as 1 arcsec to mask faint pixels with low S/N ratio further out.

In this work, the fitting procedure is achieved through the software \textsc{lenstronomy} \citep{lenstronomy1,lenstronomy2}. To reduce the complexity of model, we fix the central positions of both SIE and shear ($x_{\rm l}$ and $y_{\rm l}$ in the table) to the observable light center of the lens galaxy and only fit the other 12 parameters as listed in Tab.\,\ref{tab:model_param} through the MCMC sampling with the software \textsc{emcee} \citep{emcee_paper}. The best-fit model is the one that maximizes the likelihood function given by Eq.\,\ref{eq:likelihood}.

\begin{table*}
	\centering
    \caption{Parameters in analytical models used for lens fitting. The angular coordinates of lens $x_{\rm{l}}$ and $y_{\rm{l}}$ are fixed as the light center.}
    \label{tab:model_param}
	\begin{tabular}{lccl} 
		\toprule
		Component & Parameter & Unit & Caption\\
		\hline
        \multirow{4}{*}{Lens} & $x_{\rm{l}}, y_{\rm{l}}$ (fixed) & arcsec & Central position of lens \\
                            & $e_{\rm{l,1}}, e_{\rm{l,2}}$ & - & Eccentricity of SIE \\
                            & $\theta_{\rm{E}}$ & arcsec & Scaled Einstein radius of SIE\\
                            & $\gamma_1, \gamma_2$ & - & Shear amplitude \\
        \hline
		\multirow{5}{*}{Source} & $x_{\rm{s}}, y_{\rm{s}}$ & arcsec & Central position of source  \\
                            & $e_{\rm{s,1}}, e_{\rm{s,2}}$ & - & Eccentricity \\
                            & $n_{\rm{s}}$ & - & Sérsic index \\
                            & $R_{\rm{s}}$ & arcsec & Effective radius \\
                            & $I_0$ & $10^{-16}\ \rm{erg/s/cm^2}$ & Surface brightness at $R_{\rm{s}}$ \\
		\bottomrule      

	\end{tabular}
\end{table*}

\subsection{Source Reconstruction}
\label{sec3.3:src_rec}

Due to the effect of differential magnification, different parts of the background source are magnified differently, which results in biases into spectral analysis of source galaxy. In order to recover the intrinsic properties of the source, we present two different reconstruction methods below. The first method is using the average magnification factor based on the best-fit lens model to uniformly correct the amplitude of the overall lensed spectrum, regardless of spatial distribution. This is the case when spatially resolved spectroscopic data are not available but only a single composite spectrum can be obtained for the lensing system (Section \ref{sec3.3.1:mu_correct}). The second method is to reconstruct the spatial distribution of the source spectrum by ray-tracing from the image plane back to the source plane given the best-fit lens model. This is the case when spatially resolved spectroscopic observations for the lensed images exist (Section \ref{sec3.3.2:ray-tracing_correct}).   

\subsubsection{Correction with average magnification factor}
\label{sec3.3.1:mu_correct}
In the case that only the single composite spectrum of lensed image is available, we reconstruct the source spectrum with the average magnification factor $\bar{\mu}$ defined in Eq.\,\ref{eq:ave_mag}, with both $I(\vec{\theta})$ and $I^{\rm s}(\vec{\beta})$ obtained from the best-fit models. Then the reconstructed source composite spectrum $\mathcal{S}^{\rm s}(\lambda)$ through this method is given by $\mathcal{S}^{\rm s}(\lambda)=\mathcal{S}^{\rm o}(\lambda)/\bar{\mu}$, where $\mathcal{S}^{\rm o}(\lambda)$ is the lensed composite spectrum from observation. 

\subsubsection{Correction through full ray-tracing}
\label{sec3.3.2:ray-tracing_correct}

For the spatially resolved spectroscopic observations, we can accomplish a more delicate source reconstruction. In this work, we take the mock IFU observation of lensed image as an example and use full ray-tracing method to reconstruct source spectrum. As discussed in Section \ref{sec3.1: mock lensing systems}, lens mass distribution can be used to derive source positions $\{\vec{\beta}(\vec{\theta})\}$ corresponding to image plane positions $\{\vec{\theta}\}$ through Eq.\,\ref{eq:lenseq}. Then with the best-fit lens model, spatially resolved spectrum distribution in the source plane can be reconstructed as $\mathcal{S}^{\rm s}(\vec{\beta}(\vec{\theta}), \lambda)=\mathcal{S}^{\rm o}(\vec{\theta}, \lambda)$, where $\mathcal{S}^{\rm o}(\vec{\theta}, \lambda)$ is given by the mock IFU datacube of lensed image.

The reconstruction method described here is the inverse process of generating mock images mentioned in Section \ref{sec3.1.3:raytracing method}. To do so, we implement a pixelated mesh in the source plane, which has the same resolution as the redshift-rescaled MaNGA datacube mentioned before. We note that ray-tracing through Eq.\,\ref{eq:lenseq} can only start from the image plane to the source plane. However, due to the limited spatial resolution of the mock HST-like image and the mock IFU datacube, full ray-tracing will result in too sparse sampling of the source plane. To solve this problem, we uniformly segment every image plane pixel of the mock observation (i.e., HST-resolution for the images and MaNGA-resolution for the IFU datacube) into $5\times 5$ sub-pixels while maintaining their total spectrum flux same as the original pixel. The ray-tracing procedure is then conducted from this refined image mesh.
We note that strong lensing may produce multiple images. Therefore, an individual source plane grid at $\vec{\beta}$ in the extremely magnified regions may relate to $N$ image pixels $\{\vec{\theta_i}(\vec{\beta})\}$. (Artificial enhancement of image plane resolution can aggravate this problem.) In this case, we simply take an average among all correlated image plane grids as the reconstructed source surface brightness for that source grid point, i.e., $\mathcal{S}^{\rm s}(\vec{\beta}, \lambda)=\frac{1}{N}\sum_{i=1}^N\mathcal{S}^{\rm o}(\vec{\theta_i}(\vec{\beta}), \lambda)$.


\section{Lensed Spectra and the inferred source properties}
\label{sec4:composite spectrum after lensed}

\begin{figure*}\centering
	\includegraphics[width=\textwidth]{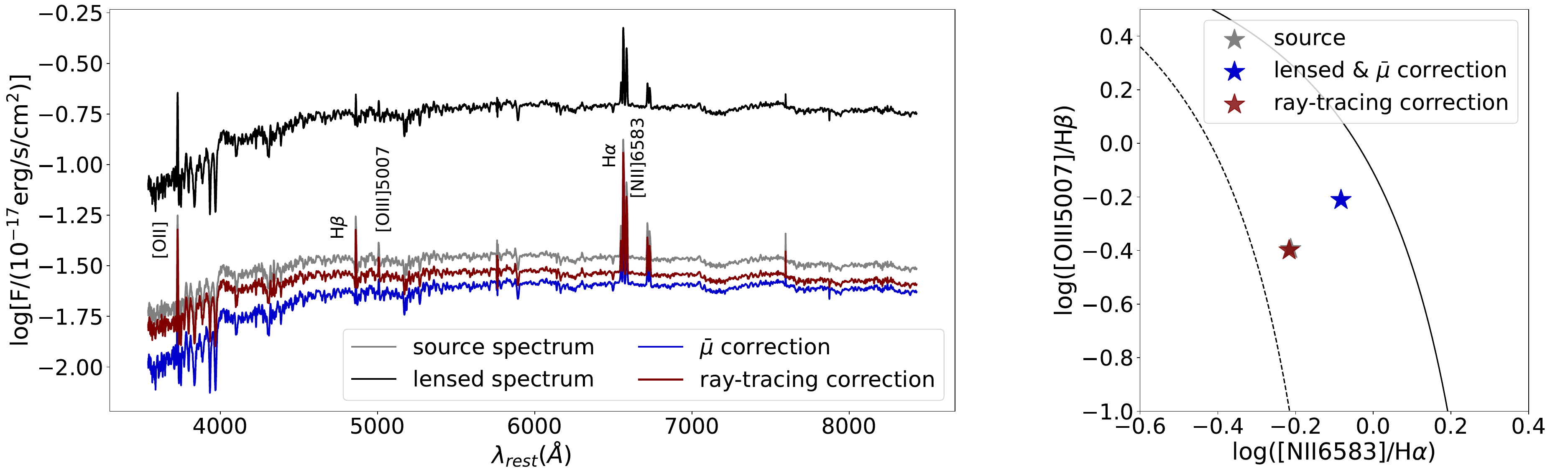}
    \caption{\textit{Left Panel:} Composite spectrum of source galaxy before lensed (gray), after lensed (black), corrected with average magnification factor $\bar{\mu}$ (blue) and corrected with ray-tracing method (red). Several emission lines are annotated in the figure. The vertical axis is adapted into logarithmic values for clarity. The horizontal axis shows the rest-frame wavelength. \textit{Right Panel:} Locations in the N2-BPT diagram of different spectra in the left panel. As discussed in Section \ref{sec5.1:correct_particular_system}, emission line ratios do not change after correction with the average magnification factor. Thus the spectra after being lensed and corrected with average magnification factor share the same location (blue star) in the BPT diagram. The gray and red stars are overlapped.}
    \label{Fig5_spec_BPT}
\end{figure*}

\begin{table*}
	\centering
    \caption{Spectral properties of source composite spectrum before lensed, after lensed, corrected with average magnification factor $\bar{\mu}$ and corrected with ray-tracing method. The values are derived with \textsc{ppxf}.}
    \label{tab:fitted_param}
	\begin{tabular}{lccc}
		\toprule
		 & $f_{\rm{H\alpha}}\ (\rm{10^{-17}erg/s/cm^2}$) & $[Z_*/H]$ & $\log(\rm{Age/yr})$\\
		\hline
        Source Spectrum & 0.78 & -0.06 & 9.10 \\
        Lensed Spectrum & 2.24 & 0.05 & 9.30 \\
        Corrected with $\bar{\mu}$ & 0.29 & 0.05 & 9.30 \\
        Corrected with Ray-Tracing& 0.67 & -0.07 & 9.09 \\
		\bottomrule      

	\end{tabular}
\end{table*}

In this section, we study the biases from lensing differential magnification effect by comparing the composite spectrum of source galaxy before and after being lensed. We focus on four spectral properties: flux of H$\alpha$ emission line, stellar metallicity, stellar age and location in the N2-BPT diagram. These properties are derived using the lensed IFU-datacube with software \textsc{ppxf}. In Section \ref{sec4.1:particular system}, we present a detailed study of the mock lensing system shown in Fig.\,\ref{Fig3_lensed_image} to investigate how the differential magnification effect biases source composite spectrum. Then in Section \ref{sec4.2:statistics}, we conduct a statistical analysis among the 400 mock systems in Section \ref{sec3.1.5:systems_in_this_work} to show the variation of such biases with different lensing configurations.

\subsection{A Particular Mock System}
\label{sec4.1:particular system}

For the mock system in Fig.\,\ref{Fig3_lensed_image}, the source galaxy is a face-on disc galaxy with a luminous central bulge, while the foreground lens is an elliptical galaxy with roundish radial and asteroid tangential caustics shown in the left panel. Fig.\,\ref{Fig5_spec_BPT} shows the composite spectrum and N2-BPT diagram of the source galaxy before and after being lensed. The spectral properties from spectrum fitting are listed in Tab.\ref{tab:fitted_param}. Comparing the two spectra shows that after being lensed, (1) flux of the continuum and H$\alpha$ line are significantly enhanced; (2) both stellar metallicity and age increase; (3) location in BPT-diagram moves towards AGN region. 

Among these changes in source spectral properties, the enhancement of continuum and Ha flux consists with the integral magnification effect of strong lensing while biases in other properties are induced by the differential magnification. 
To interpret such biases, for the luminosity-weighted quantities like stellar metallicity and stellar age, the biased values after being lensed can be represented as
\begin{equation}
\label{eq:lum_weighted_quantities}
    \bar{X}_{\rm{lensed}} \approx \frac{\sum_i \bar{\mu}_i L_i \bar{X}_i}{\sum_i \bar{\mu}_i L_i},
\end{equation}
where $\bar{\mu}_i$, $L_i$ and $X_i$ are the average magnification factor, luminosity and quantity $X$ of the $i$-th pixel in the source plane, respectively. We leave the elaboration on this equation to Appendix \ref{app:lum_weight_props}. Therefore, regions with higher magnification factors will be dominant in the weighted average. As shown in the left panel of Fig.\,\ref{Fig3_lensed_image}, in this mock system, the central bulge of source galaxy is close to the tangential caustic. Consequently, this component is magnified with a higher factor (see Section \ref{sec3.1.1: strong lensing formalism}) and dominates the composite source spectrum after being lensed. Meanwhile, Fig.\,\ref{Fig2_src_image} indicates AGN-like features and higher metallicity and age at the bulge component, which consists with the biases presented here.

\subsection{Statistics with Varied Source Positions}
\label{sec4.2:statistics}

Results in Section \ref{sec4.1:particular system} indicate that differential magnification effect can lead to marked biases in source spectral analysis. In this section, we present a statistical study among 400 mock systems with the same source and lens galaxies but different source positions to show the variations of such biases with different lensing configurations.

Fig.\,\ref{Fig6_MC_lensed} and Fig.\,\ref{Fig7_MC_BPT} present the spectral properties of the source galaxy after being lensed in each system. The points in the figures show the positions of source light center in each mock system as in Fig.\,\ref{Fig4_400systems}, but their colors now represent the changes of quantities with respect to the original source composite spectrum $\Delta X$.

As presented in the figure, the flux of H$\alpha$ emission line significantly increases after being lensed in all systems.
The stellar metallicity and age show similar biases that the values increase for source positions close to the tangential caustics and decrease for source positions further away. This consists with the negative radial gradients of stellar metallicity and age within the source galaxy and the increase(decrease) of values result from a larger magnification factor towards the central (outskirt) region. 
We further check the sources located at top left, which have a significant decrease in stellar metallicity but an increase in age after lensed. In these systems, the source galaxies are outside the tangential caustic and the differential magnification now comes from the radial caustic. As shown in Fig.\,\ref{Fig2_src_image}, region of high stellar age within the source galaxy is more extended than that of higher metallicity. As a result, differential magnification can result in a decrease in metallicity and an increase in age simultaneously when the radial caustic overlaps with regions with low stellar metallicity but high stellar age. We present more detail about such configurations in Appendix \ref{app:special_config}.

Fig.\,\ref{Fig7_MC_BPT} shows the N2-BPT diagram for each mock system and the yellow star indicates the original location of source galaxy. In line with the tight coupling between star-forming regions and H$\alpha$ emission, systems showing similar properties with star-forming galaxies in BPT diagram also have a higher lensed H$\alpha$ flux as in Fig.\,\ref{Fig6_MC_lensed}. On the contrary, due to the existence of AGN, the source becomes AGN-like when the central part is magnified with a higher factor.

The statistical results in this section indicate that the biases from differential magnification effect are prevalent and vary with different lensing configurations significantly. Compared with the always increasing H$\alpha$ flux, other properties like stellar metallicity, age and emission line ratios can have wide distributions around the true value after being lensed. In the next section, we try two different approaches to correct such biases to derive the intrinsic spectral properties of source galaxy.

\begin{figure*}
	\centering
	\subfloat[H$\alpha$ Flux]
    {
        \includegraphics[width=0.33\linewidth]{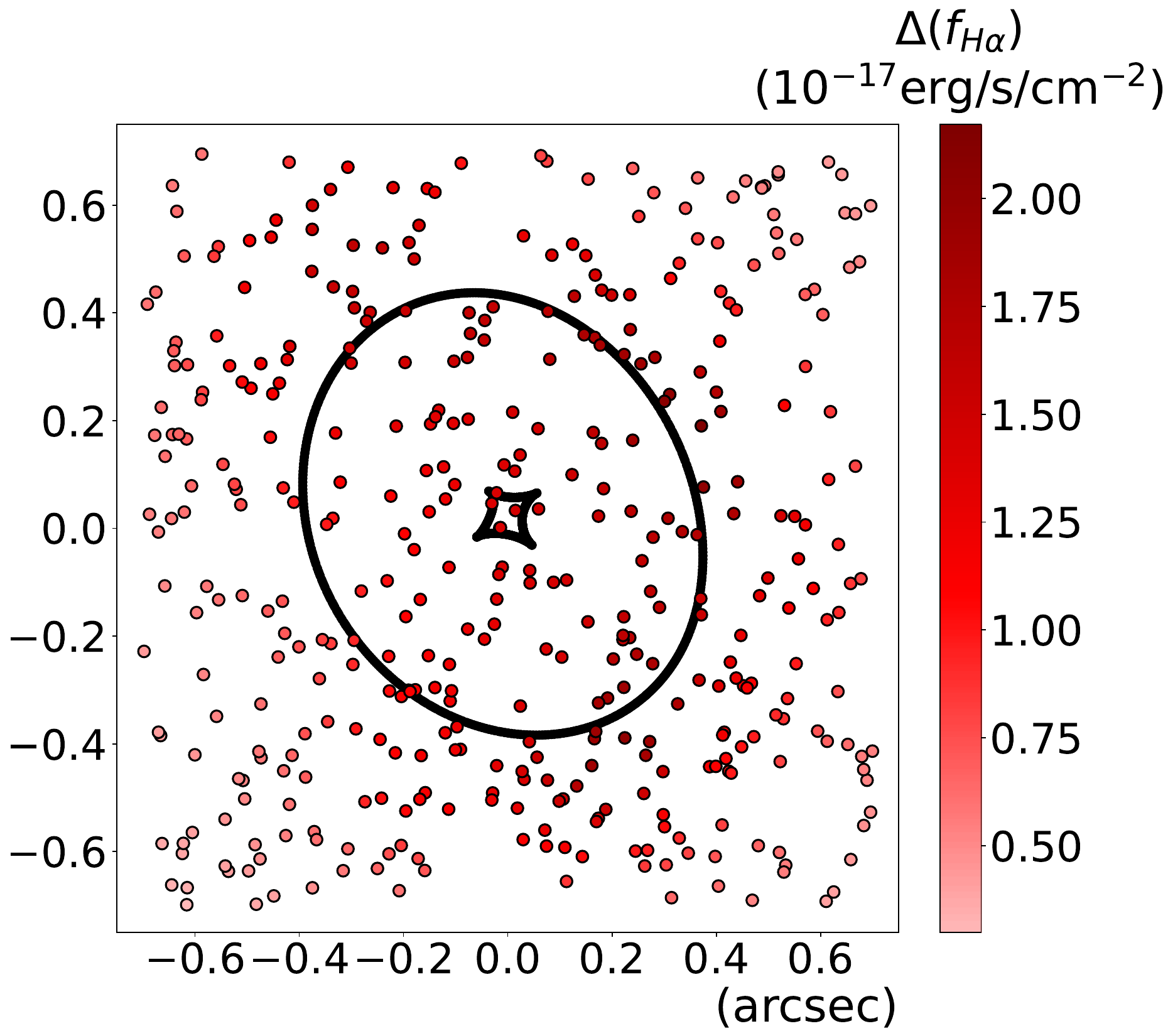}
    }
    \subfloat[Stellar Metallicity]
    {
        \includegraphics[width=0.33\linewidth]{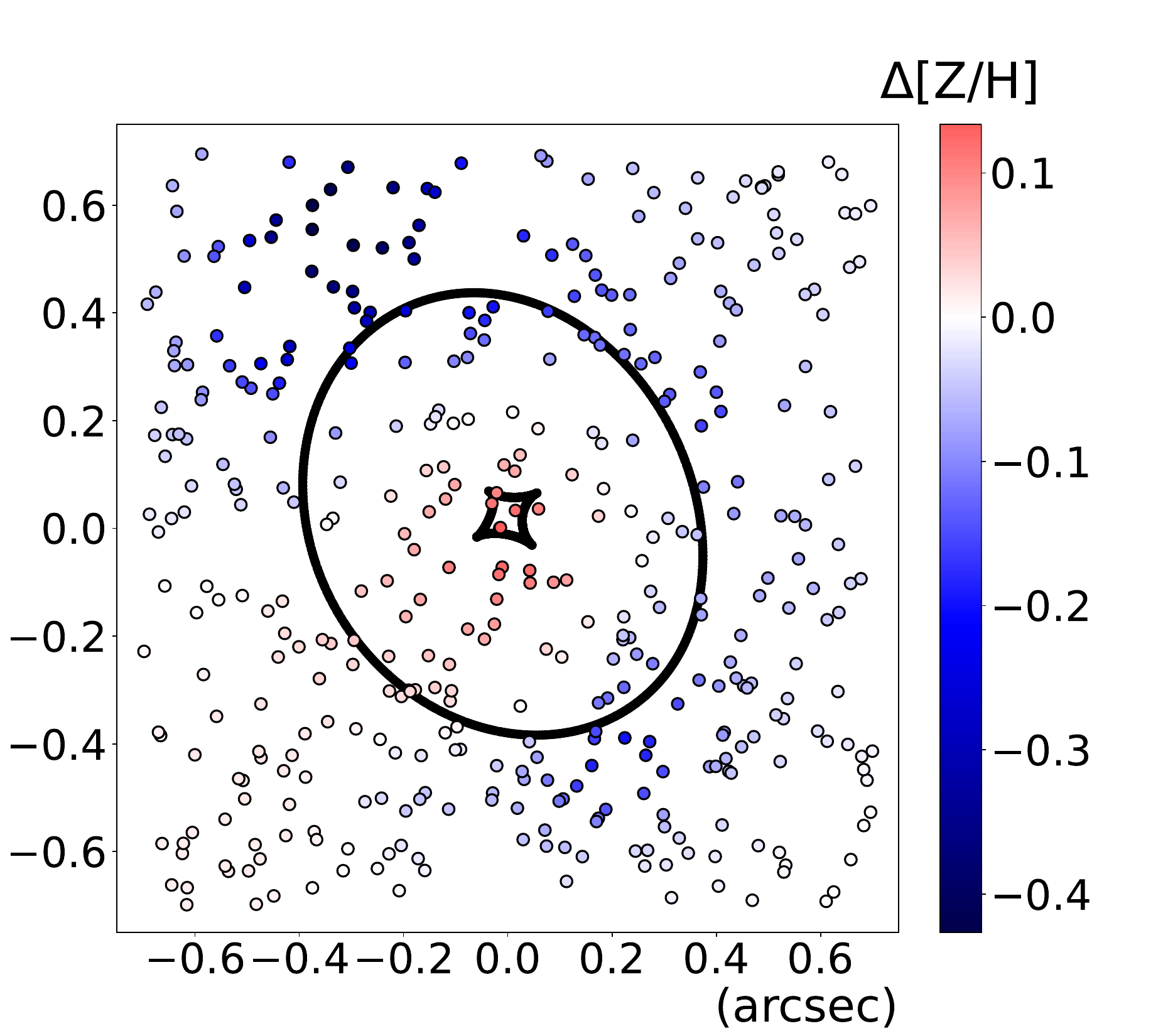}
    }
    \subfloat[Stellar Age]
    {
        \includegraphics[width=0.33\linewidth]{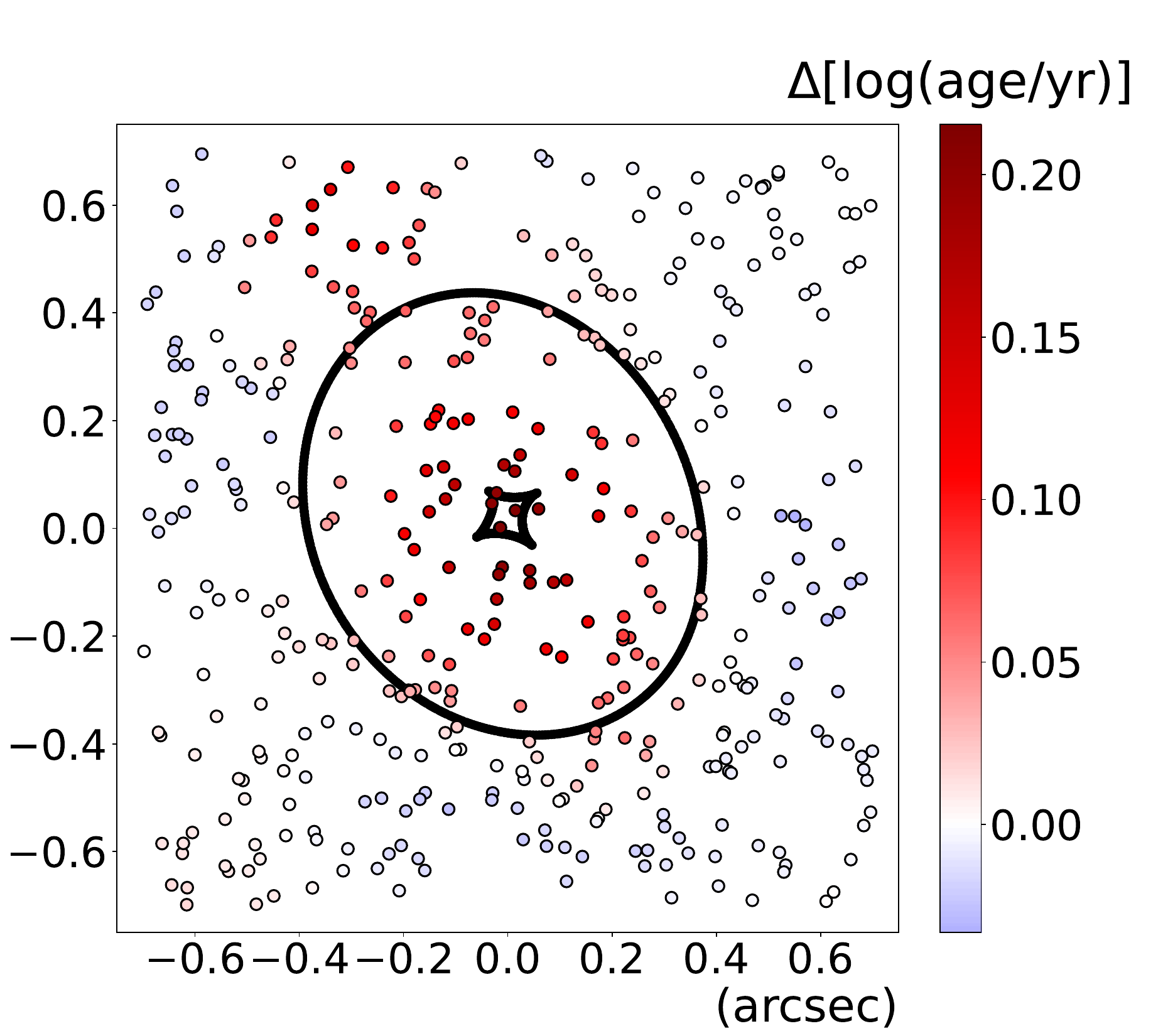}
    }

	\caption{Biased source spectral properties of the 400 mock systems in Section \ref{sec4.2:statistics}. The points indicate the positions of source light center in each system and are colour coded by the biases of different spectral properties from lensing effect ($\Delta X = X_{\rm{lensed}}-X_{\rm{source}}$).}
	\label{Fig6_MC_lensed}
\end{figure*}

\begin{figure*}\centering
	\includegraphics[width=0.67\textwidth]{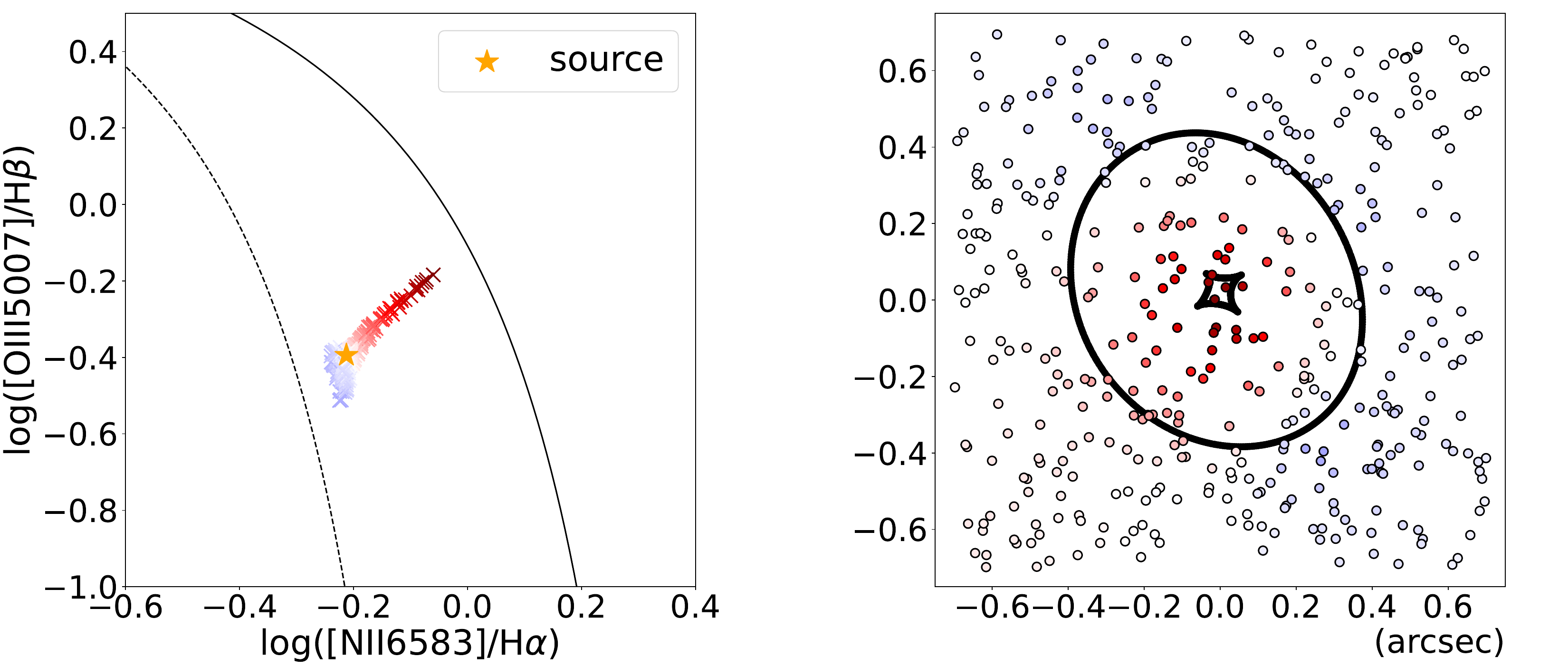}
    \caption{(Continued from Fig.\,\ref{Fig6_MC_lensed}) N2-BPT diagram of the source galaxy after being lensed in the mock systems. The colour code is used to match the crosses and points one-to-one in the two panels. The yellow star in the left panel indicates the location of the original source composite spectrum.}
    \label{Fig7_MC_BPT}
\end{figure*}


\section{Corrected intrinsic source properties}
\label{sec5:correction for biases}

To correct biases from strong lensing effect in source spectral analysis, we apply the two approaches discussed in Section \ref{sec3.3:src_rec} to our mock lensing systems and compare their capabilities. Similarly, here we focus on a particular mock system in Section \ref{sec5.1:correct_particular_system} and then present a statistical investigation in Section \ref{sec5.2:correction_statistics}.

\subsection{A Particular Mock System}
\label{sec5.1:correct_particular_system}

\begin{figure*}\centering
	\includegraphics[width=\textwidth]{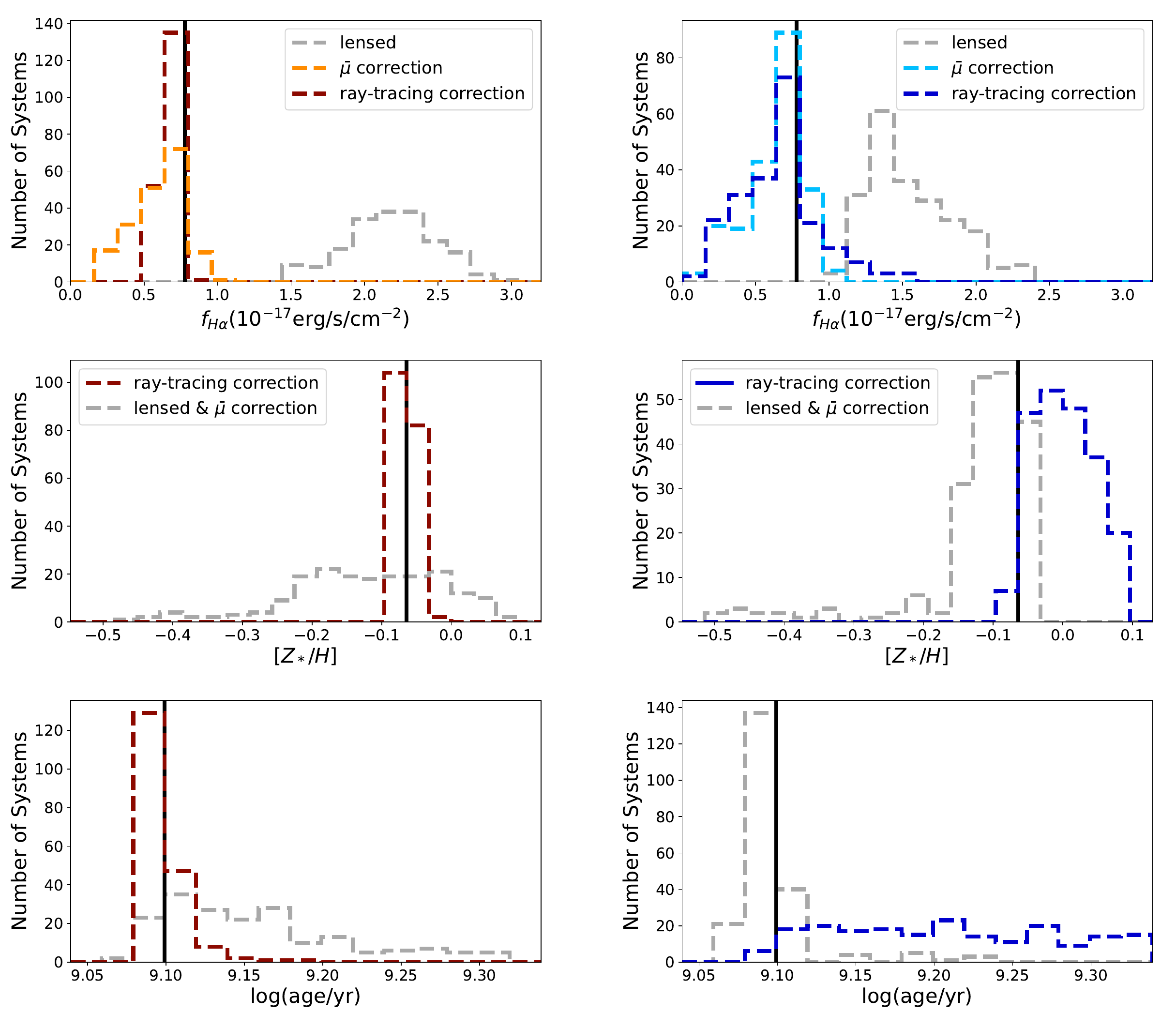}
    \caption{Correction of spectral properties with average magnification ($\bar{\mu}$) and ray-tracing method for `close’ (left column) and `far’ (right column) systems in Fig.\,\ref{Fig4_400systems}. The gray histograms show the distribution of quantities after being lensed and the red/blue ones show the correction results of ray-tracing method. For the average magnification factor method, the distribution of corrected H$\alpha$ flux is coloured in orange/cyan in the panels of the first row, while the stellar metallicity and age do not change after correction (see the context). The vertical black lines in each panel indicate the original spectral properties of the source galaxy.}
    \label{Fig8_Rec_MC_props}
\end{figure*}

In this section, we apply the two correction methods to the mock system in Fig.\,\ref{Fig3_lensed_image}. We present the fitting result of this system from MCMC sampling in Appendix \ref{app:fitting_result}. The average magnification factor calculated from Eq.\,\ref{eq:ave_mag} with the best-fit model is $\bar{\mu}\sim 7.7$ .

Fig.\,\ref{Fig5_spec_BPT} shows the composite spectrum corrected using the two methods respectively and the N2-BPT diagram. Values of other spectral properties are listed in Tab.\,\ref{tab:fitted_param}. Comparing the two reconstructed source spectra shows that (1) the corrected flux of the continuum and the  H$\alpha$ emission lines are underestimated in both methods but the average magnification factor gives a larger discrepancy; (2) other spectral properties like stellar population properties and emission line {\it ratios} cannot be corrected with average magnification factor, while the ray-tracing method works well.

It is obvious that the average magnification factor from the best-fit model is inadequate to correct biases from differential magnification effect and can involve extra biases, such as the underestimation of H$\alpha$ flux shown in Tab.\,\ref{tab:fitted_param}. The key reason for this is that the average magnification factor here is derived from a single-band image. Since galaxies hold complicated and spatially varying spectral properties, the differential magnification effect indicates that the average magnification factor in the form of Eq.\,\ref{eq:ave_mag} can significantly change with wavelength. In real observations, the average factor used for source reconstruction is estimated from images in exclusively one waveband (or several particular wavebands) and such dependency on wavelength is incapable of modeling. As a result, the properties corrected in this way can still be biased. For instance, the ratio of emission lines, which changes after their being magnified differently, remains invariant after correcting their flux with the same average factor derived from the single-band image. The same things happen to stellar metallicity and age (Tab.\ref{tab:fitted_param}), which depend on the equivalent width of absorption lines and shape of the continuum, respectively.

Instead, ray-tracing method performs much better in recovering all of the spectral properties studied here. The effectiveness of ray-tracing method relies on the achromaticity of lensing deflection effect, which means that the deflecting behaviour of lens object is the same for light rays of any wavelength. 
Following this property, although lensing modeling here only utilizes the V-band image, the derived deflection angles $\vec{\alpha}(\vec{\theta})$ from the best-fit lens model can be used for ray-tracing in all wavebands. Therefore, ray-tracing method can recover spectral properties of the background source precisely in theory. We note that the remaining biases in the source properties corrected in this way come from the inaccuracy of parametric models used in lens modeling and the observational effects in the image plane IFU data, such as the spatial resolution.

\subsection{Statistical Study of Two Correction Methods}
\label{sec5.2:correction_statistics}

\begin{figure*}\centering
	\includegraphics[width=0.8\textwidth]{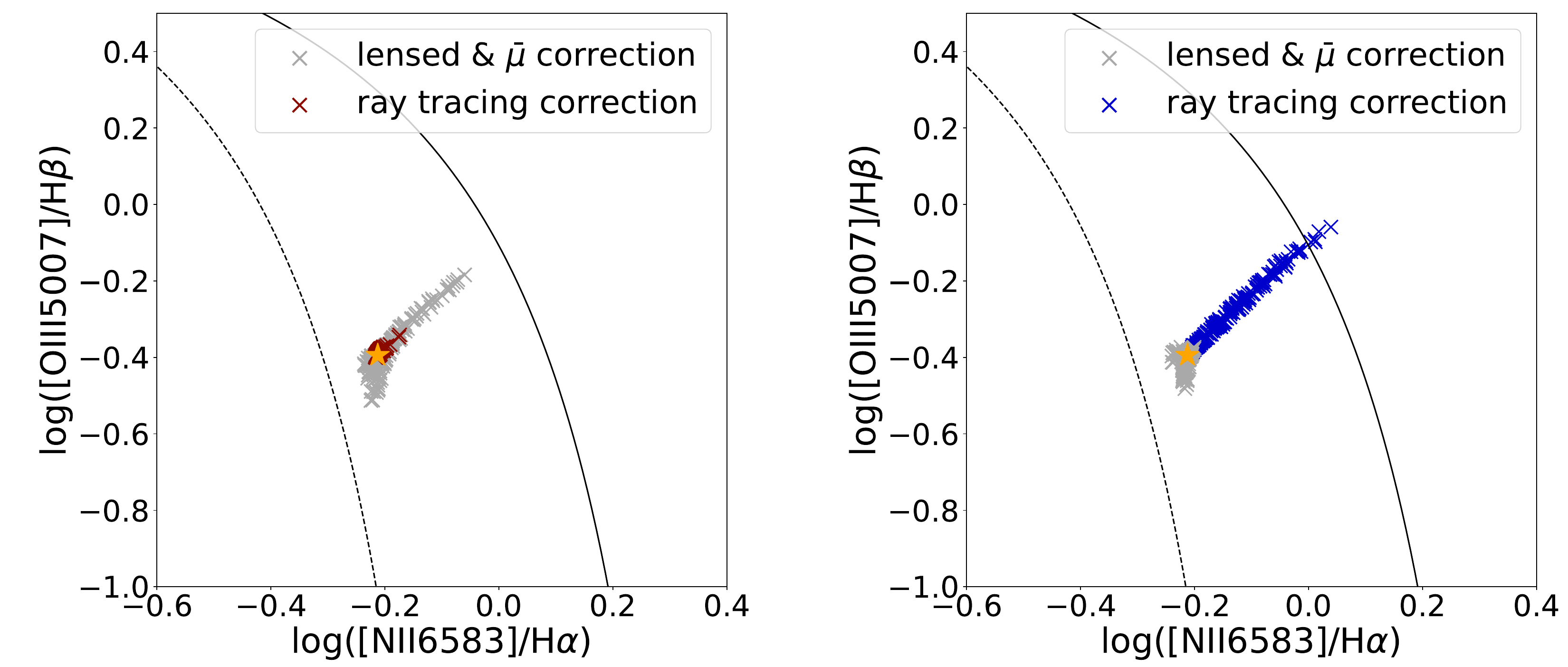}
    \caption{(Continued from Fig.\,\ref{Fig8_Rec_MC_props}) N2-BPT diagram of source galaxy corrected with the two methods for `close’ (left panel) and `far’ (right panel) systems, respectively. The yellow stars indicate the location of the original source composite spectrum.}
    \label{Fig9_Rec_MC_BPT}
\end{figure*}

In this section, we further apply the two methods to the suite of mock systems to compare their capabilities from the statistical perspective. Since our lens modeling in this work only utilizes the lensed image (see Section \ref{sec3.2:lens_modeling}), the constraint on lens model becomes tighter as more and more components of source galaxy are strongly lensed into multiple images. Therefore, we further divide the 400 systems into `close’ and `far’ subgroups as discussed in Section \ref{sec3.1.5:systems_in_this_work}.

Comparison between the two groups of mock systems in Fig.\,\ref{Fig8_Rec_MC_props} and Fig.\,\ref{Fig9_Rec_MC_BPT} shows that the biases are much stronger in the `close’ systems, which consists with the fact that a larger part of the source galaxy is close to the caustics and strongly lensed in these systems. We find the same result as Section \ref{sec5.1:correct_particular_system} that spectral properties like stellar metallicity, age and emission line ratios cannot be corrected using average magnification factor. In the `close’ systems, ray-tracing method works much better than the average magnification factor in correcting all of the source spectral properties studied here. However, the performance of ray-tracing method become much worse in the `far’ systems and it also involves extra biases in the corrected source spectral properties, such as stellar age and emission line ratios as shown in the figures. 

We attribute the poor performance of ray-tracing method in the `far’ systems to the inaccuracy of best-fit lens model. From simulation data, we have verified that the density profile of the selected lens galaxy can be well described by an SIE model. Owing to different source positions, in `close' systems the central bulge in source galaxy forms luminous arc-shaped images after being lensed and imposes tight constraints on lens model, while only the fainter disc gets strongly lensed for sources further away. As a result, the lens model for the `far’ systems are less constrained and significantly affected by parameter degeneracies, resulting in extra biases in subsequent correction procedures. The comparison between `far’ and `close’ systems here indicates the importance of accurate lens modeling to the faithful correction of biases. Therefore, when dealing with strong lensing systems in real observations, it is necessary to include more constraints, such as imaging or dynamics of the foreground lens galaxy, into lens modeling to break the degeneracy between different parameters.


\section{Discussion and Summary}
\label{sec6:summary}

Magnification of strong lensing effect varies with different source positions. For an extended source galaxy, different components are magnified with different factors and such differential magnification effect can markedly bias the properties of source composite spectrum. In this work, we generate mock strong lensing systems using selected MaNGA galaxy as source and IllustrisTNG galaxy as lens, to study such biases in source spectral analysis. We have then examined two different correction methods for such biases and compare their effectiveness. 

Our main findings are listed here:

\begin{enumerate}
    \item Composite spectral properties of source galaxy change significantly after being strongly lensed. Strong lensing magnification effect enhances flux of emission lines while differential magnification involves biases into source spectral properties, such as stellar metallicity, age and emission line ratios (N2-BPT diagram). Such biases vary with different strong lensing configurations.
    
    \item Correction with average magnification factor, which is generally employed in observations, is incapable of calibrating the biases from differential magnification. As different parts of the source exhibiting different emission signatures with wavelength get magnified differently, without spatially resolved spectroscopic observations, the average magnification factor predicted by the best-fit lens model based on single-band images cannot correct for properties that depend on emissions at different wavelengths, such as stellar population properties and emission line ratios as shown in Fig.\,\ref{Fig5_spec_BPT} and Tab.\,\ref{tab:fitted_param}. 
    
    \item Correction through ray-tracing method based on the achromatic nature of strong lensing deflection effect can effectively restore source spectral properties. However, this correction method relies on the accuracy of lens modeling, as inaccurate models introduce extra biases into the reconstructed source composite spectrum, as discussed in Section \ref{sec5.2:correction_statistics}.
\end{enumerate}

The biases induced by differential magnification effect arise from the interplay between the spatial variation of properties within source galaxies and the magnification gradient in the source plane. 
Consequently, it changes with several factors, including the properties and inclinations of the source galaxy, which can give different spectral distributions in the source plane, as well as the density distribution of lens galaxy and lensing configurations, which make source galaxies experience different magnification gradients. These factors also add complexity into lens modeling and further influence the correction of such biases. 
In this demonstrative study, we select a single MaNGA galaxy with large radial gradients of spectral properties as source, and a massive elliptical galaxy from TNG-100 as the foreground lens to generate mock systems. While we investigate the variation of biases with different lensing configurations by changing source positions, this specific choice of source and lens galaxies can represent an extreme scenario. A more comprehensive study to check other factors using more realistic mock lensing systems with different source and lens galaxies is planned for future work. Such analysis is important to the statistical study of strongly lensed high-redshift galaxies. We also note that galaxies at $z\sim2.0$ are anticipated to exhibit properties different from local galaxies in MaNGA survey. Therefore, mock IFU datacubes generated from hydrodynamic simulations, such as TNG, will be valuable for following investigations.

Both of the two correction methods discussed here rely on the best-fit model from lens modeling procedure. In this study, we exclude substructures within the selected lens galaxy and external shear from the large-scale structure, which results in smooth caustics as shown in Fig.\,\ref{Fig3_lensed_image}. 
However, the structure and surrounding environment of lens galaxies can be far more complex in real observations. Therefore, it is necessary to improve the complexity of parametric models or cross-check between different modeling strategies when applying these correction methods to real lensing systems. Similar to galaxy-scale strong lensing systems, differential magnification also presents in group- and cluster-scale systems, the modeling of which is inherently much more complicated. However, since sources are typically lensed into multiple arclet images in these systems, it is possible to derive the properties of source galaxies from images located far from the critical lines, which is less biased by the differential magnification effect \citep{Motta2018}.

Although we focus on the properties derived from lensed spectrum in this work, the differential magnification effect also introduces significant biases into the spectral energy distribution (SED) data that are more available for most strong lensing systems, as already indicated in previous studies \citep[e.g.][]{Blain1999, Zotti2024}. As a result, properties of source galaxies from SED fitting can be biased when using uncorrected SED data or those corrected with the average magnification factor derived in a single waveband. Additionally, it is important to note here that the lack of detailed corrections can also lead to systematic biases in photometric redshifts, which utilize SED data \citep[e.g.][]{photoz}.

Current and future sky survey projects, such as the ground-based \textit{Vera C. Rubin Observatory} \citep[LSST,][]{LSST_survey} and space-borne \textit{China Space Station Telescope} \citep[CSST,][]{CSST_survey}, \textit{Euclid} \citep{Euclid_survey} and the \textit{Nancy Grace Roman Space Telescope} \citep[Roman,][]{Roman_survey}, are predicted to increase the number of strong lensing systems by several orders of magnitude. Follow-up spatially resolved spectroscopic observations, such as slitless spectroscopy and IFS, can provide valuable observations to faithfully recover intrinsic spectral properties of background source galaxies. 
The investigation in this work has been achieved through assuming HST-like image observations and MaNGA IFU-like spectroscopic data. However, future investigations along this line implementing other observational conditions, such as MUSE at the ground-based \textit{Very Large Telescope} \citep[VLT/MUSE,][]{MUSE} and the IFS aboard the advanced \textit{James Webb Space Telescope} \citep[JWST,][]{Boker2022, Rigby2023}, can be easily achieved.


\section*{Acknowledgements}

We acknowledge the very helpful discussions with Drs. Shengdong Lu and Xiaoyue Cao. This work is supported by the China Manned Space Project (No. CMS-CSST-2021-A07), the General Program from National Natural Science
Foundation of China (No. 12073013), the Beijing Municipal Natural Science Foundation (No. 1242032), and the National Key Research
and Development Program of China (No. 2023YFA1607904). JG acknowledges support from the Youth Innovation Promotion Association of the Chinese Academy of Sciences (No. 2022056). We also acknowledge the high-performance computing cluster in the Astronomy Department of Tsinghua University.


\section*{Data Availability}

In this work, we use data from the MaNGA survey and IllustrisTNG project to generate mock strong lensing systems. The MaNGA DRP and DAP datacubes are publicly available in \url{https://www.sdss4.org/dr17/manga/manga-data/data-access/}. The MaNGA DynPop data we use to select source galaxies are available in \url{https://manga-dynpop.github.io}. The data of TNG-simulations can be obtained in \url{https://www.tng-project.org}.


\bibliographystyle{mnras}
\bibliography{reference} 



\appendix

\section{Lens Modeling Results}
\label{app:fitting_result}

\begin{figure*}\centering
	\includegraphics[width=\textwidth]{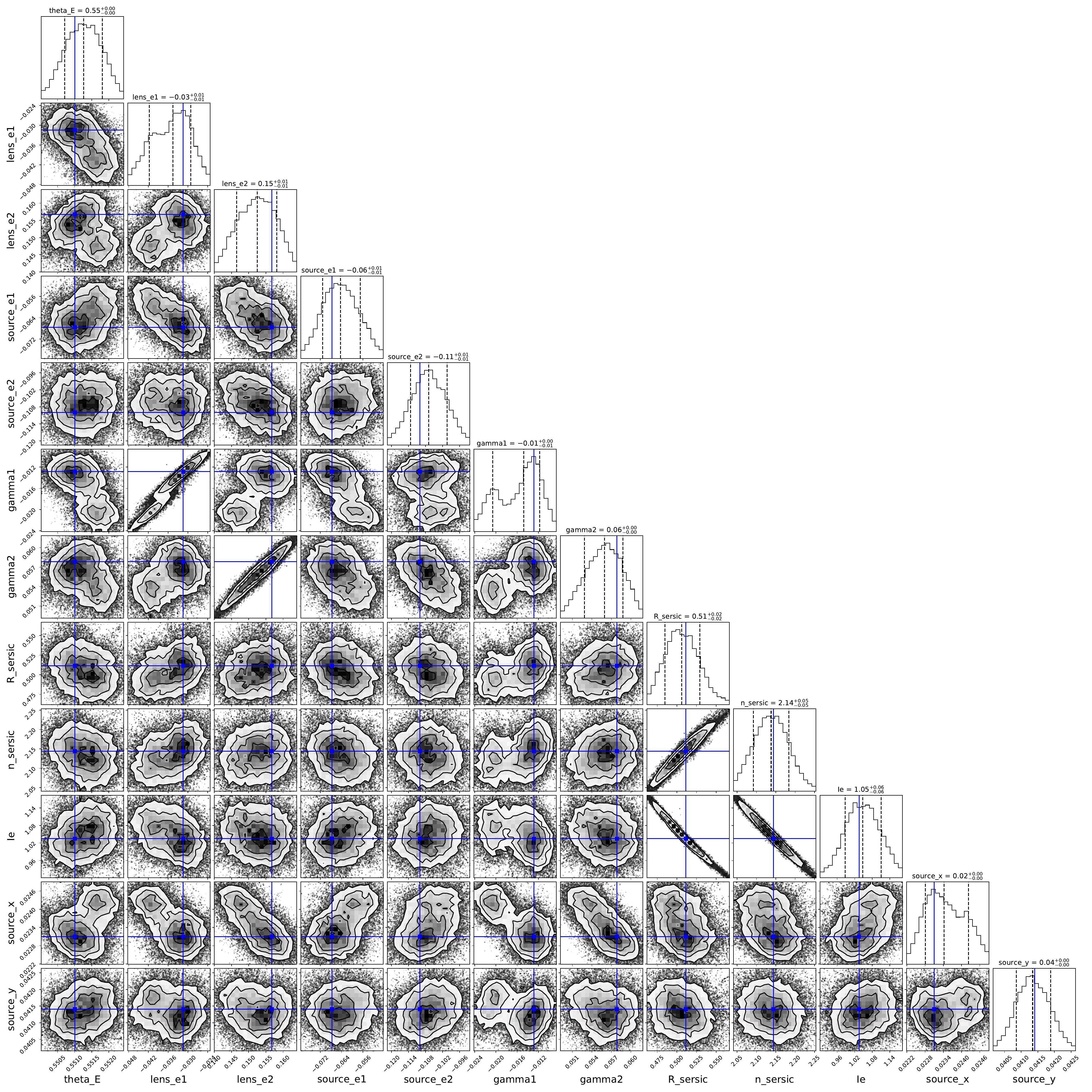}
    \caption{Corner plot showing the results of MCMC. The blue lines here indicate the best-fit parameters maximizing the likelihood.}
    \label{figA1:corner_plot_MCMC}
\end{figure*}

In the lens modeling process, we apply the MCMC method to maximize the likelihood with the software \textsc{emcee}. We use 120 chains (10 times of the number of free parameters) and 5000 steps with the first 2000 steps for burn-in. It is noteworthy that around 15\% of chains fail to converge and we discard them manually. Fig.\,\ref{figA1:corner_plot_MCMC} shows the posterior probability distribution, which indicates strong degeneracy among several particular parameters. The bimodal marginalized distributions of several parameters indicate a local maximum of the likelihood. We attribute it to the complexity of likelihood in lens modeling with extended images and the limitations of parametric models. As mentioned in the main text, we take the set of parameters maximizing the likelihood as the best-fit parameters, which are shown with blue lines in the figure. The best-fit values of all parameters are within the 1$\sigma$ regions of the marginalized distributions.

\begin{figure*}\centering
	\includegraphics[width=0.8\textwidth]{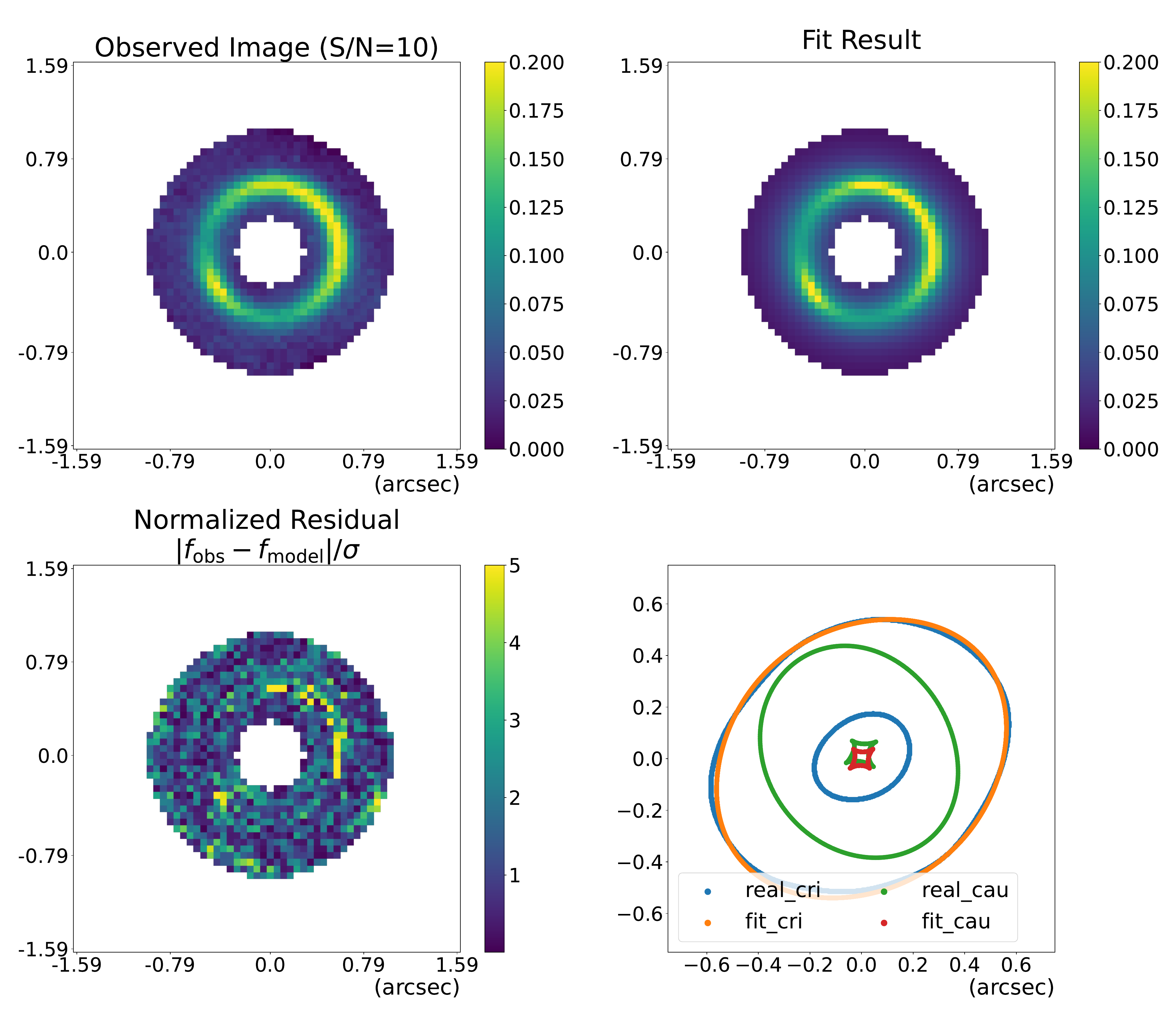}
    \caption{Comparison of the observed image and the image generated from the best-fit model of the mock system in Fig.\,\ref{Fig3_lensed_image}. The bottom two panels show the normalized residual map and the caustics and critical curves of the best-fit model.}
    \label{figA2:comparison}
\end{figure*}

Fig.\,\ref{figA2:comparison} shows the comparison between the input image and image generated from the best-fit model. Here we also compare the caustics and critical curves of the best-fit model with the simulated lens galaxy. Due to the singularity of the analytical SIE model employed here, the best-fit model has no radial caustics and critical curves as shown in the figure.

\section{A special configuration}
\label{app:special_config}

\begin{figure*}\centering
    \includegraphics[width=\textwidth]{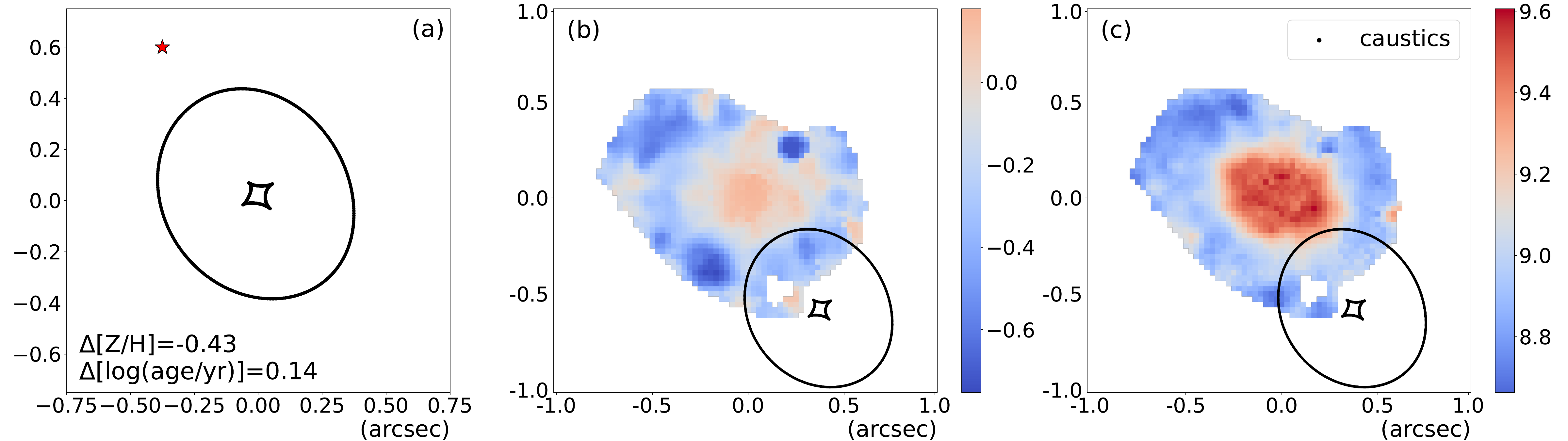}
	\caption{Relative position of source galaxy and caustics in the mock system discussed in Appendix \ref{app:special_config}. Panel (a) shows the source light center with a star with the biases of stellar metallicity and age annotated aside. Panel (b) and (c) show the stellar metallicity and age maps overlapped with the caustics, respectively.}
	\label{FigB_special_config}
\end{figure*}

Here we study the mock systems with higher stellar metallicity but lower stellar age after being lensed, which locate at the top left in Fig.\,\ref{Fig6_MC_lensed}, in more detail. We take one mock lensing system as an instance here, as the relative position of source and caustics is shown in Fig.\,\ref{FigB_special_config}. As the tangential caustic move outside the source galaxy, the radial caustic now indicates regions with a high magnification factor. 

Comparing Panel (b) and Panel (c) indicates that, due to the greater extension of regions with high metallicity than those of old stellar population, source regions close to the radial caustics in this system show low stellar metallicity but still a high stellar age. As a result, after being magnified, the biased source stellar metallicity becomes higher while the stellar age rather decreases. 

\section{Luminosity-weighted properties after being lensed}
\label{app:lum_weight_props}

In this work, we take a spiral galaxy as source galaxy for case study and use the \textsc{ppxf} package \citep{Cappellari2004,Cappellari2017} for spectral fitting. Considering that spiral galaxies have simpler star formation histories than post-starburst galaxies and their luminosity- and mass-weighted stellar population parameters are close to each other \citep{Ge2021}, in this paper, we take the luminosity-weighted stellar population properties for analyses. In this case, the input spectra are fitted with spectral templates of multiple stellar populations and the resulting spectral properties are weighted by the luminosity of these populations
\begin{equation}
\label{eq:app_ave_prop}
    \bar{X}=\frac{\sum_i L_i X_i}{\sum_i L_i},
\end{equation}
where $X_i$ and $L_i$ are the spectral properties and luminosity of the $i$-th template, and the summation goes over all templates used for fitting.

When we derive the composite spectral properties from pixelated IFU datacubes, the quantities can be represented as
\begin{equation}
\label{eq:app_c2}
\begin{split}
    \bar{X}_{\rm comp}&=\frac{\sum_i \left(\left(\sum_J L_{i, J}\right) X_i\right)}{\sum_i \left(\sum_J L_{i, J}\right)}=\frac{\sum_J \left(\sum_i \left(L_{i, J} X_i\right)\right)}{\sum_J \left(\sum_i L_{i, J}\right)}\\
    &=\frac{\sum_J \left(\bar{X}_J \left(\sum_i L_{i, J} \right)\right)}{\sum_J \left(\sum_i L_{i, J}\right)}=\frac{\sum_J L_J \bar{X}_J}{\sum_J L_J},
\end{split}
\end{equation}
where $i$ refers to the $i$-th stellar population template and $J$ represents the $J$-th pixel in the datacube. $\bar{X}_J$ is the properties of $J$-th pixel derived from Eq.\,\ref{eq:app_ave_prop}. This indicates that the composite spectral properties of galaxy can also be represented as the luminosity-weighted average of the properties of different components.

For a strongly lensed source galaxy, the area of source galaxies is magnified and differential magnification effect introduces spatially varying magnification factors. The magnification of area increases the total number of summation in Eq.\,\ref{eq:app_c2}, and equivalently introduces another weight into the average
\begin{equation}
\label{eq:app_c3}
    \bar{X}_{\rm{comp,\,lensed}}= \frac{\sum_J \bar{\mu}_J L_J \bar{X}_J}{\sum_J \bar{\mu}_J L_J},
\end{equation}
where $J$ now refers to the $J$-th pixel in the source plane and $\bar{\mu}_J$ is the average magnification factor of the $J$-th pixel as defined in Eq.\,\ref{eq:ave_mag}.

It is noteworthy that the discrete summation among pixelated datacubes here is an approximation of the continuous integral over the whole galaxy. Nevertheless, Eq.\,\ref{eq:app_ave_prop}--\ref{eq:app_c3} with integrals should be in the same form.


\bsp	
\label{lastpage}
\end{document}